# Intervalence Plasmons in Boron-Doped Diamond

Souvik Bhattacharya,[1,*] Jonathan Boyd,[2,*] Sven Reichardt,[3,*], Valentin Allard,[4] Amir Hossein Talebi,[3] Nicolò Maccaferri,[5] Olga Shenderova,[6] Aude L. Lereu,[4] Ludger Wirtz,[3] Giuseppe Strangi,[2,**] and R. Mohan Sankaran[1,**]

[1]Department of Nuclear, Plasma, and Radiological Engineering, The Grainger College of Engineering, University of Illinois Urbana-Champaign, Champaign, IL, U.S.A.

[2]Department of Physics, Case Western Reserve University, Cleveland, OH, U.S.A.

[3]Department of Physics and Materials Science, University of Luxembourg, Luxembourg

[4]Aix Marseille Univ, CNRS, Centrale Méditérranée, Institut Fresnel, Marseille, France

[5]Department of Physics, Umeå University, Sweden

[6]Adamas Nanotechnologies, Raleigh, NC, U.S.A.

*These authors contributed equally to this work.

**Corresponding authors: gxs284@case.edu and rmohan@illinois.edu

## Abstract

Doped semiconductors can exhibit metallic-like properties ranging from superconductivity to tunable localized surface plasmon resonances. Diamond is a wide-bandgap semiconductor that is rendered electronically active by incorporating a hole dopant, boron. While the effects of boron doping on the electronic band structure of diamond are well-studied, any link between charge carriers and plasmons, has never been shown. Here, we report intervalence plasmons in boron-doped diamond, defined as collective electronic excitations between the valence subbands, opened up by the presence of holes. Evidence for these low-energy excitations is provided by valence electron energy loss spectroscopy and near-field infrared spectroscopy. The measured spectra are subsequently reproduced by first-principles calculations based on the contribution of intervalence band transitions to the dielectric function. Our calculations also reveal that the real part of the dielectric function exhibits a crossover characteristic of metallicity. These results suggest a new mechanism for inducing plasmon-like behavior in doped semiconductors, and the possibility of attaining such properties in diamond, a key emerging material for quantum information technologies.

## Introduction

Doping of semiconductors by the introduction of impurities imparts novel electrical[1,2] and optical[3,4] properties that discretely depend on the dopant level. In lightly-doped semiconductors, impurities are isolated down to as small as a single atom, and optically addressable electronic and spin states are produced.[5-7] In heavily-doped semiconductors, the increased charge carrier concentration leads to metal-like behavior.[8,9] Localized surface plasmon resonances have been observed,[10,11] but unlike metals, the resonant frequency is tunable over an order of magnitude or more by varying the charge carrier concentration.[12,13]

Among semiconductor materials, diamond holds a special place, as it is endowed with exceptional transmissibility from ultraviolet to microwave frequencies, thermal conductivity, breakdown field, and charge carrier mobility at room temperature.[14] Doping to relatively high concentrations can be achieved with a charge acceptor, boron, which has a small atomic radius and is thus highly soluble in the relatively dense diamond lattice.[15] At very high concentrations ($>10^{20}$ cm$^{-3}$), metallic properties have been observed in boron-doped diamond (BDD), including the emergence of superconductivity.[16] The band dispersion has been measured by angle-resolved photoemission spectroscopy and the metallicity has been linked to holes in the intrinsic diamond bands.[17] However, other effects of hole doping beyond electronic conductivity remain unexplored. In particular, the opening of intervalence band (IVB) transitions, as reported for other hole-doped semiconductors,[18,19] and theorized to be responsible for changes in phonon frequencies[20], should leave a direct imprint on the electronic excitation spectrum.

In this work, we employ scanning transmission electron microscope-valence electron loss spectroscopy (STEM-VEELS) and near-field infrared (IR) spectroscopy to probe low-energy excitations in BDD. We develop a first-principles model to calculate the measured inelastic loss and optical absorption from the dielectric function. Our results reveal that the origin of the material response, which is absent in undoped diamond, is IVB transitions created by hole doping, and the response is plasmonic in character.

## Results

**Electronic band structure of BDD.** Figure 1a presents the band structure of intrinsic diamond near the center of the first Brillouin zone. With a bandgap of ~5.5 eV, there are effectively no free charge carriers in intrinsic diamond at room temperature and any electrical conductivity that has been measured is the result of electrochemically-mediated surface transfer doping.[21] Doping diamond with single boron impurities results in an acceptor level approximately 0.37 eV above the top of the valence band (Fig. 1b).[22] Electrons at the top of the valence band are able to fill the relatively low-energy acceptor states, creating holes in the valence band. With increasing boron concentration, more and more electrons depopulate the valence band, leaving behind more and more holes deeper and deeper within the valence band. In p-type semiconductors like BDD, the formation of holes in the valence band opens up intervalence (IVB) transitions between degenerate subbands, specifically the light-hole (LH) and heavy-hole (HH) bands.

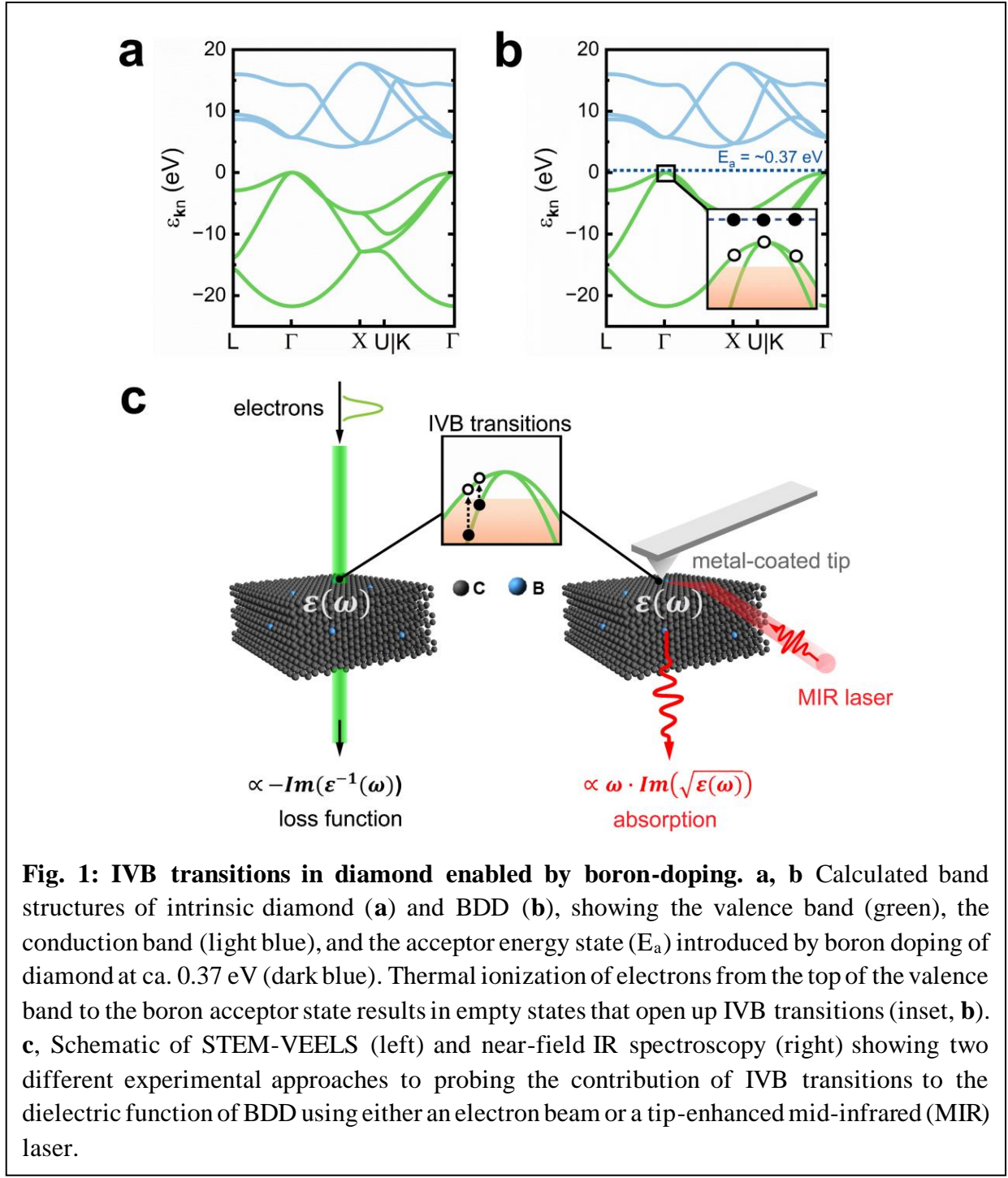

**Fig. 1: IVB transitions in diamond enabled by boron-doping. a, b** Calculated band structures of intrinsic diamond (**a**) and BDD (**b**), showing the valence band (green), the conduction band (light blue), and the acceptor energy state ($E_a$) introduced by boron doping of diamond at ca. 0.37 eV (dark blue). Thermal ionization of electrons from the top of the valence band to the boron acceptor state results in empty states that open up IVB transitions (inset, **b**). **c**, Schematic of STEM-VEELS (left) and near-field IR spectroscopy (right) showing two different experimental approaches to probing the contribution of IVB transitions to the dielectric function of BDD using either an electron beam or a tip-enhanced mid-infrared (MIR) laser.

For this study, BDD samples were synthesized by the high-pressure, high-temperature (HPHT) method and as observed by transmission electron microscopy (TEM), were found to consist of highly crystalline particles ~1 μm in diameter, shaped like "broken glass" because of the grinding process after synthesis (Supplementary Figs. 1 and 2). As a reference, nearly identical undoped HPHT diamond particles were also synthesized by the same procedure (Supplementary Figs. 1 and 2). The boron concentration in the BDD was below the detection limit of core-loss EELS (Supplementary Note 1 and Supplementary Figs. 2 and 3).

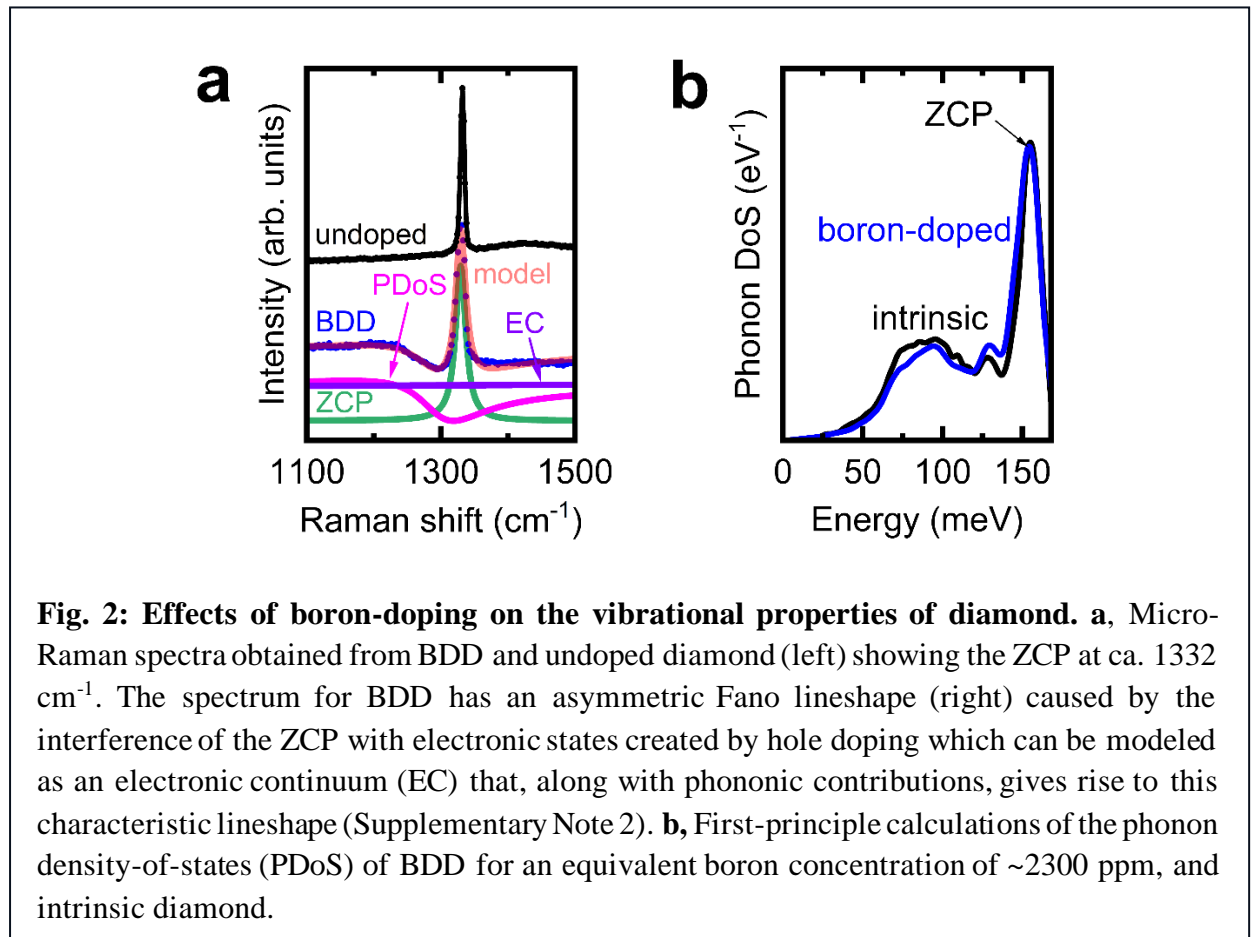

**Fig. 2: Effects of boron-doping on the vibrational properties of diamond. a**, Micro-Raman spectra obtained from BDD and undoped diamond (left) showing the ZCP at ca. 1332 cm$^{-1}$. The spectrum for BDD has an asymmetric Fano lineshape (right) caused by the interference of the ZCP with electronic states created by hole doping which can be modeled as an electronic continuum (EC) that, along with phononic contributions, gives rise to this characteristic lineshape (Supplementary Note 2). **b,** First-principle calculations of the phonon density-of-states (PDoS) of BDD for an equivalent boron concentration of ~2300 ppm, and intrinsic diamond.

**Phononic properties of BDD.** We initially characterized the BDD by micro-Raman spectroscopy which shows the well-known zone center phonon (ZCP) of diamond at ca. 1332 cm$^{-1}$ (Figure 2a). In comparison to undoped diamond, boron doping is found to produce an asymmetry in the Raman line shape of the ZCP, commonly attributed to the Fano resonance effect, i.e., an interference between the Raman-active phonon and a continuum of electronic states. The observation of a Fano resonance in BDD provides a first hint of the modification of the electronic structure by the free hole population. Indeed, the nature of the Fano-type line shape in BDD is sensitive to the boron concentration, and was fit based on a previously reported model to determine a hole density of ~$5 \times 10^{19}$ cm$^{-3}$ in our sample (Supplementary Note 2).[23] We note that with a laser spot size of 10 µm, the boron concentration obtained by micro Raman spectroscopy represents an average over several particles. In order to probe variations in the boron concentration not only particle to particle, but potentially within a particle, tip-enhanced Raman spectroscopy (TERS) was carried out, which overcomes the diffraction limit of micro-Raman spectroscopy. By extending our analysis of the Fano-induced asymmetry in the ZCP peak, we found that the boron concentration varied between particles and even within particles from $3 \times 10^{19}$ cm$^{-3}$ to $1.8 \times 10^{20}$ cm$^{-3}$ (Supplementary Fig. 4). The fundamental origin of the electronic transitions in BDD that result in the Fano effect has not yet been explained by theory or probed by experiments, but in boron-doped silicon has been correlated to IVB transitions whose energy overlaps with that of the ZCP.[24] In addition to the asymmetric Raman lineshape, the ZCP peak is also slightly red shifted, which has been attributed to the interaction of the phonon with IVB transitions.[20] This negligible change of the ZCP is also seen by comparing first-principles simulations of the full phonon density (PDOS) of BDD and undoped diamond (Figure 2b).

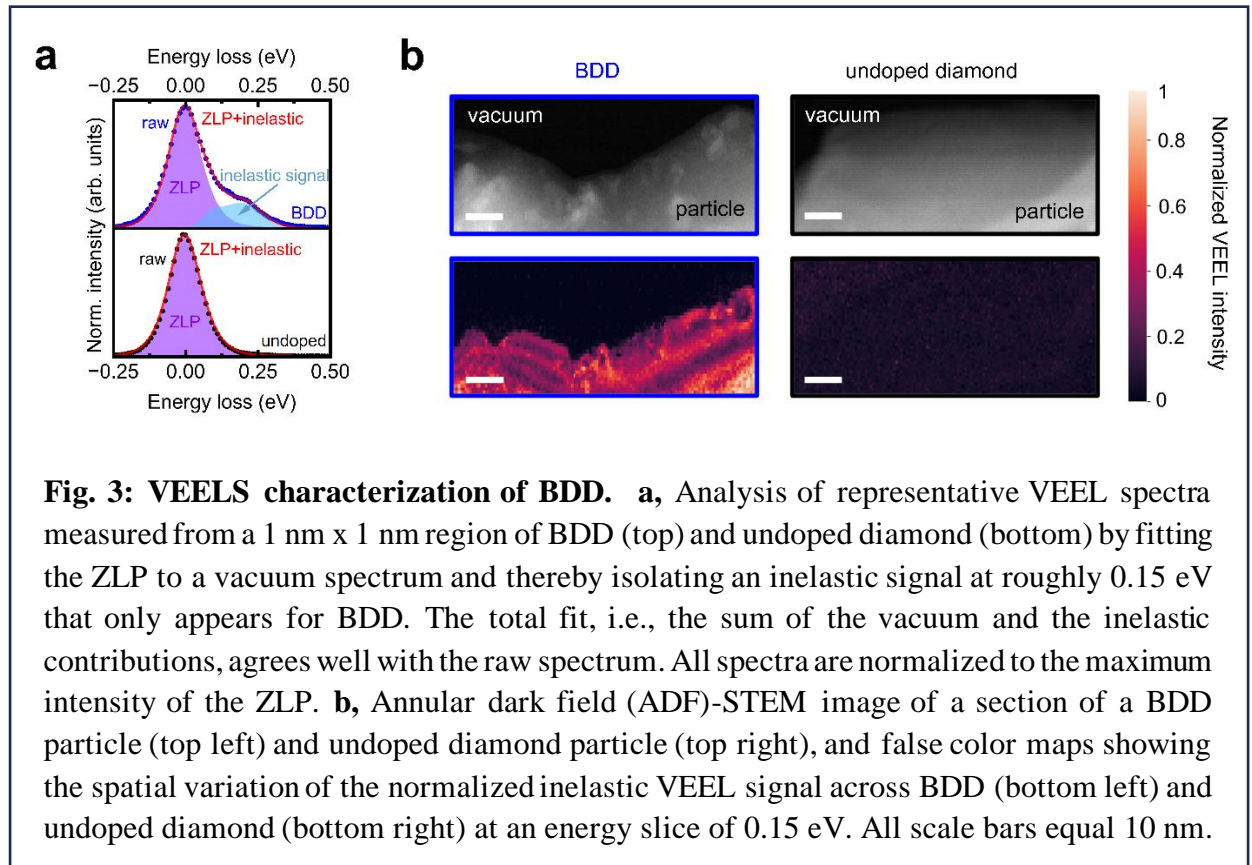

**Fig. 3: VEELS characterization of BDD. a,** Analysis of representative VEEL spectra measured from a 1 nm x 1 nm region of BDD (top) and undoped diamond (bottom) by fitting the ZLP to a vacuum spectrum and thereby isolating an inelastic signal at roughly 0.15 eV that only appears for BDD. The total fit, i.e., the sum of the vacuum and the inelastic contributions, agrees well with the raw spectrum. All spectra are normalized to the maximum intensity of the ZLP. **b,** Annular dark field (ADF)-STEM image of a section of a BDD particle (top left) and undoped diamond particle (top right), and false color maps showing the spatial variation of the normalized inelastic VEEL signal across BDD (bottom left) and undoped diamond (bottom right) at an energy slice of 0.15 eV. All scale bars equal 10 nm.

**Experimental measurements of the dielectric response of BDD.** To further probe the electronic structure of BDD, we performed STEM-VEELS and near-field IR spectroscopy (Fig. 1c). In VEELS, fast electrons in the electron beam are inelastically scattered by a material. Recent advancements in electron microscopy, including monochromated electron beams, high-resolution electron energy-loss spectrometers, and a narrow zero-loss peak tail, have made it possible to analyze energy losses down to 10 meV.[25] In comparison, near-field IR spectroscopy measures the optical response of materials. Here, we employ two variations of near-field IR spectroscopy: 1) photo-induced force microscopy (PiFM) and 2) scattering scanning near-field optical microscopy (s-SNOM), which are both techniques broadly based on atomic force microscopy and focusing a laser onto the tip to create a local field that confines light-matter interaction at the nanoscale. While PiFM utilizes the resultant localized mechanical deflection of the tip to measure optical absorbance,[26,27] s-SNOM measures the optical amplitude and phase shift of the scattered field, which can then be demodulated at higher harmonics to extract the near-field optical interaction.[28-30] In addition, these techniques allow for a sample's optical response to be mapped on a length scale far exceeding the diffraction limit of light. Together, STEM-VEELS and near-field IR spectroscopy provide nanometer-scale spatial resolution in the low-energy range necessary to probe intervalence band transitions (<0.5 eV), while complementing each other in that the combination of the two yields a complete view of longitudinal and transverse excitations, respectively.

Representative VEELS measurements for BDD and undoped diamond are shown in Fig. 3a. The spectrum for BDD consists of an elastic zero loss peak (ZLP) and a clearly observable shoulder, corresponding to an inelastic contribution between ~0.1 and 0.3 eV. The inelastic signal was isolated by fitting the ZLP to a vacuum spectrum and fitting a Voigt profile (Supplementary Note 4 and Supplementary Fig. 5). Similar spectra were detected and analyzed across different regions of a given particle, as well as from many different BDD particles (Supplementary Fig. 6). In comparison, no inelastic signal is present in the representative spectrum for undoped diamond (see Fig. 3a) or any of the additional spectra that were collected across a given particle surface and different particles (Supplementary Fig. 7).

While only the BDD produced an observable inelastic VEELS signal, we applied the same fitting procedure to all spectral analysis of undoped diamond as well (Supplementary Figs. 5 and 7). The lack of any signal other than the ZLP peak is consistent with the large band gap of undoped diamond, which suppresses any electronic contribution. Since our VEELS measurements were carried out at the optical limit, i.e., with the momentum transfer or scattering vector, $q_\perp = 0$, phonon modes are not probed for a nonpolar material like diamond.[31,32] As optical phonons can be excited in both doped and undoped diamond, as shown by Raman spectroscopy, we conclude that the VEELS inelastic signal is not phononic in nature. In addition, this signal at <0.5 eV is clearly distinct from the volume plasmon peak found at ~34 eV, which is present in both the BDD and the undoped samples (Supplementary Fig. 2) and originates from collective electronic excitations of electrons from the valence to the conduction band.

To study the spatial dependence of the inelastic signal measured by VEELS, we collected and analyzed spectra over a particle area. The signal was affected by the thickness of the particle and was higher near the edge where the particle is thinner. Using the log-ratio method, we estimate that the thickness varies from ~2-3 nm near the edge, to ~15-20 nm at the thickest part of the particle that was characterized (Supplementary Note 4).[33] Nonetheless, we were able to obtain signal ~tens of nm away from the edge and normalize the measurements by referencing with the ZLP, which provided a measure of the overall signal attenuation. Figure 3b shows a STEM image and corresponding map of the inelastic signal for a 134 nm x 39 nm section of a representative BDD particle, where the color intensity corresponds to the integrated normalized intensity of the inelastic loss peak at an energy of 0.15 eV, the nominal energy value obtained from VEELS. For reference, a STEM image and map is also shown for a 120 nm x 46 nm section of a representative undoped diamond particle. The inelastic signal intensity is observed to vary across the BDD particle, exhibiting a striated pattern, but does not appear to be localized to a specific part of the particle such as the edge, or depend on the loss energy. Maps of other BDD particles showed similar random intensity distributions (Supplementary Fig. 8) and did not show any correlation with particle thickness (Supplementary Fig. 9). As expected based on VEELS spectra, the maps of undoped diamond samples did not show any appreciable intensity (Fig. 3b and Supplementary Fig. 10). These results indicate that the primary signal does not arise from defects intrinsic to the material such as at the surface, which should also be present in undoped diamond, but that boron doping is responsible. The spatial variations in the boron concentration, which was indicated by TERS analysis, could be related to the distribution of the boron atoms in the diamond lattice, which is known to prefer certain crystal directions,[34,35] or structural heterogeneities that have been recently found to impact the properties of other impurity dopants in diamond at the atomic scale.[36]

Representative PiFM spectra are shown in Fig. 4a for BDD and undoped diamond. A peak is observed for BDD centered at ca. 0.13 eV, close to the energy values measured by VEELS. Additional spectra were collected from different regions and different particles for both BDD and undoped diamond (Supplementary Fig. 11). Once again, spectral features were only found in BDD samples. In comparison to VEELS, there is no interference with the peak of interest and the broad nature is clear. Using a laser frequency close to the nominal peak observed in PiFM of 1048 cm$^{-1}$, s-SNOM imaging was conducted (see Supplementary Note 6 and Methods for details). Figs. 4b-g show representative AFM topography, optical amplitude, and optical phase images obtained from isolated crystals of BDD and undoped diamond. Similar quantitative changes are observed in the amplitude for both BDD and undoped diamond (see Figs. 4d, 4e and Supplementary Fig. 12), but importantly, a distinct and uniform contrast in the phase is observed across BDD that is completely absent in undoped diamond (see Figs. 4f, 4g, and Supplementary Fig. 12).

**Modeling the complex dielectric function of BDD.** VEELS and near-field IR spectroscopic techniques like PiFM and s-SNOM probe the frequency-dependent dielectric response of a material to an external electric field. Specifically, VEELS measures the energy dissipation, i.e., loss channels, and is represented by the negative imaginary part of the inverse dielectric function, and PiFM and s-SNOM measure quantities related to absorption, which is proportional to the imaginary part of the square root of the dielectric function.[37] In doped semiconductors, the charge carriers introduced into the valence or conduction band can enable electronic transitions beyond the ones present in the undoped case and lead to additional contributions to the complex dielectric function. Specifically, for a p-type semiconductor such as BDD, electrons in the hole-doped band may be excited within the band, so-called intravalence band transitions. At the same time, holes also allow the excitation of an electron from the filled light hole band into the hole of in the heavy hole band, so-called IVB transitions (see Fig. 1b). To understand the physical origin of the electronic transitions, we model these various modes of excitations.

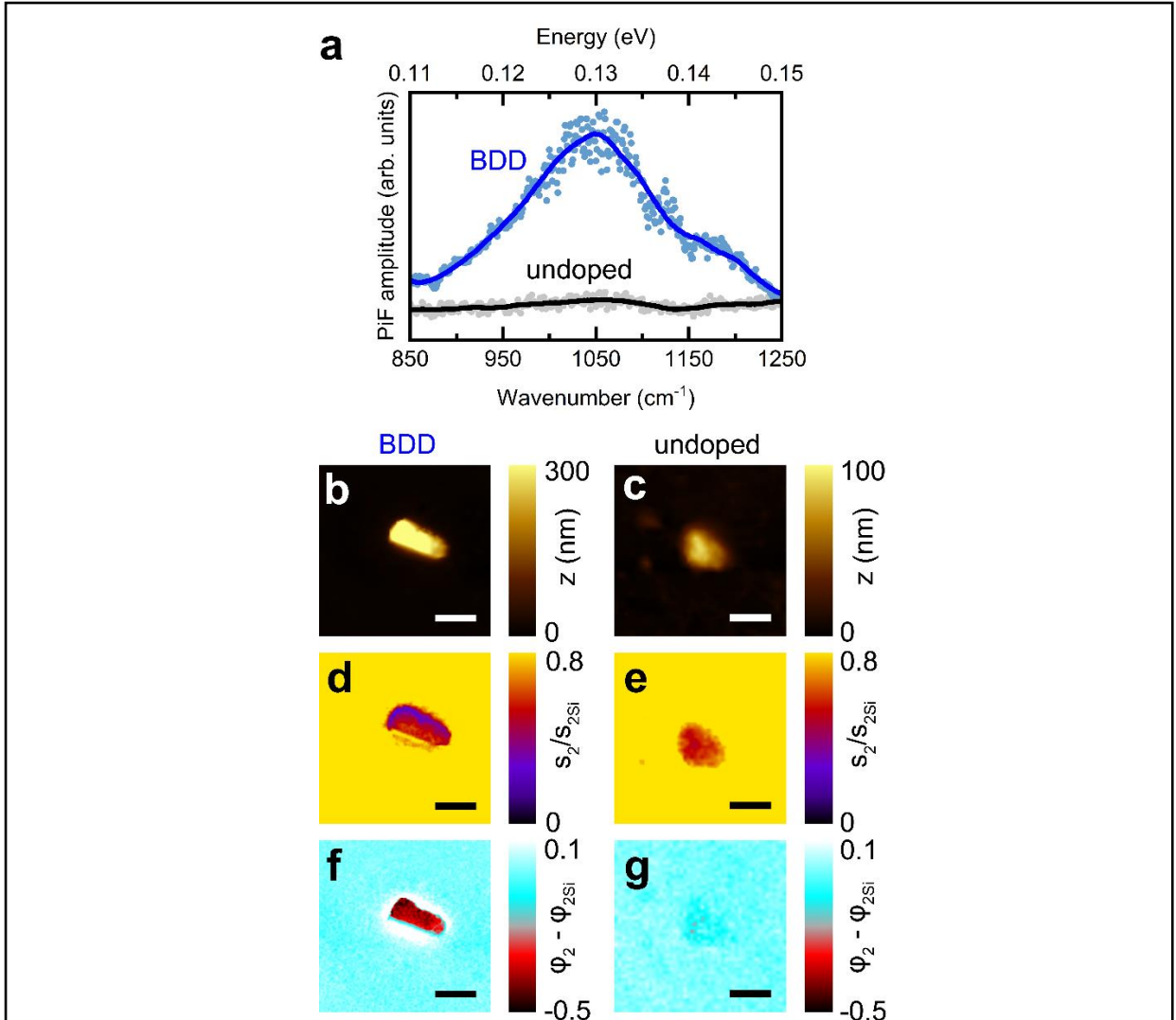

**Fig. 4: Near-field infrared spectroscopic characterization of BDD. a**, Representative photo-induced force microscopy (PiFM) spectra collected from BDD and undoped diamond. **b-g**, Representative s-SNOM images collected from BDD and undoped diamond: atomic force microscopy (AFM) topography images **(b)** and **(c)**, respectively, amplitude, $s(\omega)$, **(d)** and **(e)**, respectively, and phase, $\varphi(\omega)$, **(f)** and **(g)**, respectively. Laser energy was 1048 cm$^{-1}$ (~0.13 eV), corresponding to the resonance observed in (a). The amplitude and phase signals correspond to the second harmonic of the tip oscillation frequency and were normalized with reference to the Si substrate. All lateral scale bars correspond to 1 μm.

First, a modified version of a one-frequency Drude-Lorentz model known as the Thomas-Fermi model, which has been previously reported for hole excitations in doped semiconductors, was applied.[18,38] At a hole density of ~$5 \times 10^{19}$ cm$^{-3}$ corresponding to our BDD sample, we calculate a loss peak at 0.7 eV, which is at much higher energy than our experimentally-measured VEELS inelastic signal peak at ~0.15 eV, and indicates that the origin is not Drude-like, i.e., not due to intravalence band excitations (Supplementary Fig. 13 and Supplementary Note 5). Next, we consider IVB transitions opened up by the depopulation of electrons and formation of holes in the valence band, as illustrated in Fig. 1b. To find the complex inverse dielectric function, we performed first-principles calculations (Supplementary Note 6). At the moderate concentrations of boron studied here, we assumed that the band structure of BDD is equivalent to intrinsic diamond, with an effective Fermi level related to the boron-induced free hole density, as depicted in Fig. 5a, which has been previously shown to successfully reproduce the vibrational band structure of BDD as a function of boron concentration.[20] The Fermi energy was obtained by matching the numerically integrated occupancies of all electronic valence band states to the hole density. The resulting non-linear relation of the Fermi energy shift in BDD, $\Delta\varepsilon_F$, with the hole density is shown in Fig. 5b. We then calculated the contribution of all possible intervalence band transitions to the complex macroscopic dielectric function. The total dielectric function was determined by adding the static dielectric constant of intrinsic diamond, which accounts for additional contributions of high-energy electronic transitions from the valence to conduction bands. A static approximation for these interband contributions is justified as our analysis focuses on low excitation energies (<0.5 eV) that are much smaller than the band gap of intrinsic diamond (ca. 5.5 eV). Finally, the VEELS loss function and the IR absorption coefficient are obtained, as detailed in the Methods. We chose to model the absorption coefficient as both the signal measured by PiFM and the phase contrast detected by s-SNOM have been found to be linked to opti0cal absorption.[30,39]

The resulting simulated spectra for VEELS loss and IR absorption are shown in Fig. 5c. The spectral position of the maximum loss for both loss mechanisms are found to strongly depend on the hole density. The simulated VEELS peaks shift from 0.05 eV at $\sim 1.8 \times 10^{19}$ cm$^{-3}$ to 0.3 eV at $\sim 1.8 \times 10^{20}$ cm$^{-3}$. In comparing with experimental results, we note that the measured VEELS peak energies vary slightly from particle-to-particle and across particles in the range of 0.1-0.2 eV (Supplementary Figs. 6 and 8), which according to the simulation results shown in Fig. 5c would correspond to a range of hole densities from ~3.5 to $8 \times 10^{19}$ cm$^{-3}$, which agrees well with the variation in the boron concentrations obtained by TERS analysis. To quantify the experimental variation in VEELS, spectra were analyzed from every 1 nm x 1 nm spot of the representative particle analyzed in Fig. 3, and a distribution of peak energies was compiled and found to exhibit the most frequent occurrence at ~0.12-0.13 eV, corresponding to a hole density of $3.5 \times 10^{19}$ cm$^{-3}$, which agrees well with the average boron concentration obtained by micro Raman analysis (Supplementary Fig. 14). The simulated IR loss peaks display a similar dependence on hole density, shifting from 0.1 eV at $\sim 1.8 \times 10^{19}$ cm$^{-3}$ to 0.3 eV at $\sim 1.8 \times 10^{20}$ cm$^{-3}$. In comparing with experiment results, the measured PiFM peak energy and the resonant condition of s-SNOM at ~.13 eV would, based on simulations, correspond to a hole density of $\sim 5 \times 10^{19}$ cm$^{-3}$, which agrees well with the average boron concentration obtained by micro Raman analysis.

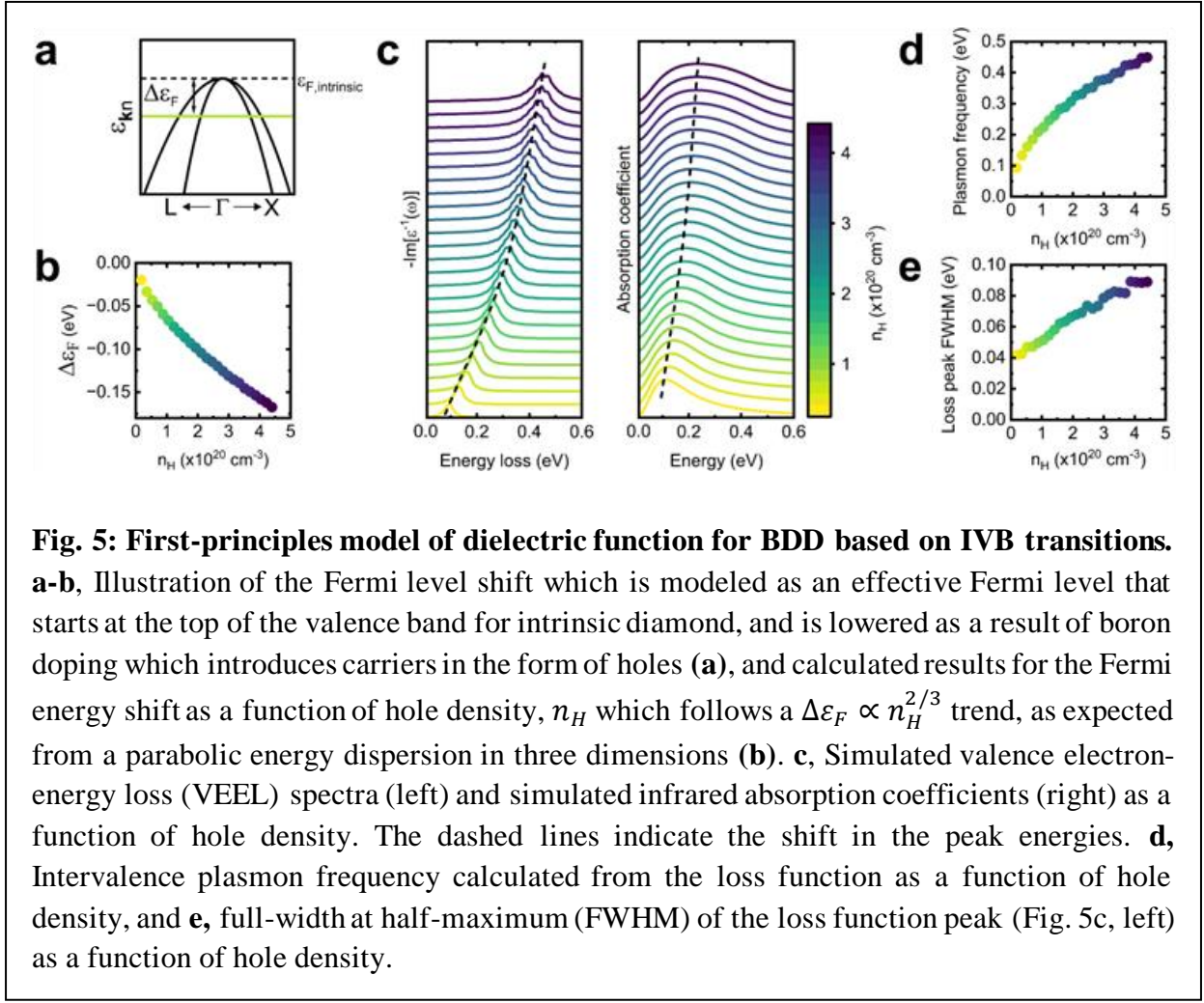

**Fig. 5: First-principles model of dielectric function for BDD based on IVB transitions.**
**a-b**, Illustration of the Fermi level shift which is modeled as an effective Fermi level that starts at the top of the valence band for intrinsic diamond, and is lowered as a result of boron doping which introduces carriers in the form of holes (**a**), and calculated results for the Fermi energy shift as a function of hole density, $n_H$ which follows a $\Delta\varepsilon_F \propto n_H^{2/3}$ trend, as expected from a parabolic energy dispersion in three dimensions (**b**). **c**, Simulated valence electron-energy loss (VEEL) spectra (left) and simulated infrared absorption coefficients (right) as a function of hole density. The dashed lines indicate the shift in the peak energies. **d,** Intervalence plasmon frequency calculated from the loss function as a function of hole density, and **e,** full-width at half-maximum (FWHM) of the loss function peak (Fig. 5c, left) as a function of hole density.

By analyzing the modeled dielectric function, we observe that the peak present in the loss spectrum corresponds to a zero-crossing (i.e., negative band crossing) of the real part of the dielectric function, indicating a clear transition to metallic properties (Supplementary Fig. 15). This result is qualitatively consistent with s-SNOM measurements where a difference was observed in the phase for BDD without a change in amplitude from undoped diamond. Furthermore, the blue shift in the frequency of this resonance (Fig. 5d) and the resulting loss peak, as well as the increase in the intensity of the loss peak with increasing hole density, are consistent with plasmonic behavior. Overall, these insights lead us to define the collective electronic transitions in the valence subbands of BDD as intervalence plasmons. Note that this type of plasmon is entirely different from both the Drude-like plasmons found in metals linked to intra-band transitions, and from the volume plasmon also found in undoped semiconductors at higher energies, which arise from valence-to-conduction band transitions. By analyzing the full-width-half-maximum (FWHM) of the loss-function peak, we find that the dampening constant which is proportional to the inverse lifetime of a plasmon, is well below the eigenfrequency, indicating that the observed intervalence plasmon is a well-defined quasi-particle (Fig. 5e). The damping increases with hole concentration, as the available phase space for the individual electron-hole pair decay channel increases, stemming from the monotonically increasing density of intervalence band transition states in diamond. However, the plasmon frequency likewise increases with hole concentration, such that the intervalence plasmon is not subject to overdamping. To further underscore the well-defined quasi-particle nature, we compare the position and FWHM of the simulated loss-function peak with a direct calculation of the plasmon frequency and lifetime from the real and imaginary part of the dielectric function (Supplementary Fig. 15) and find excellent agreement.

## Discussion

Through a combination of experimental measurements and *ab initio* modeling, we report plasmonic behavior at infrared frequencies in BDD that can be attributed to hole doping, not directly as free charge carriers, but more indirectly by enabling subband transitions for valence band electrons. This mechanism is distinct from the intravalence band excitations of free charge carriers of a particular type (e.g., holes for p-type semiconductors) that have typically been modeled by the Drude model to explain the optical response of doped semiconductors.[40] The question of whether the same models used to describe the plasmonic response from metals are applicable has recently been raised,[41] and our results provide key physical insight into how doped semiconductors are different. We note that BDD may be unique because of its wide band gap, large acceptor level energy, and low dielectric polarizability as compared to other, similar doped semiconductors. As a result, the intervalence band response may be unfettered compared to, for example, either a narrow-band gap semiconductor, or one where the dopant level is closer to the valence band. The broader implication of our study is a much richer and more complex picture for the effect of doping on electronic transitions in semiconductors, which could enable the engineering of semiconductor-based plasmonic materials with different types of band-to-band excitations.

Unlocking plasmonic properties in diamond also adds another interesting element to this remarkable material. Diamond has recently attracted attention because of the ability to host emissive point defects with quantum properties, which combined with its small spin-orbit coupling and zero nuclear spin,[42] permits the electron spin to be initialized, manipulated, and optically read-out at room temperature.[43,44] However, the emitters have been limited by long fluorescent lifetimes, low photon rates, and poor quantum efficiency, which are important for quantum applications. In general, co-doping impurities that are electronically active and color centers, respectively, could use plasmonic excitation to enhance the photoluminescence and speed up the emission rates.[45] In addition, the charge state of the point defects could be altered as a result of the shift in the Fermi level and pinning induced by the electronic dopant, such as in the case of the recently reported neutralization of $SiV^-$ to $SiV^0$, to produce new emitters with disruptive improvements in properties.[46] The discovery of these and other point defects is rapidly expanding the toolkit for engineering quantum emitters.[47] Electronic doping provides another lever to optimize the optical and spin properties. Future studies will need to focus on controlling the nanoscale concentration of dopants such as boron to tune the various effects.

## Methods

### Synthesis of diamond powders
Diamond powders were synthesized by metal-catalyzed (Fe and Ni) high-pressure, high-temperature (HPHT) methods using graphite as the carbon precursor. Boron carbide ($B_4C$) was mixed with the graphite as the boron source to obtain BDD. Both the as-synthesized undoped and boron-doped powders were washed thoroughly and treated with hydrochloric acid to eliminate metal contaminants and remove amorphous carbon.

### TEM imaging
TEM was performed using a FEI Tecnai G2 F20 microscope at an accelerating voltage of 200 kV. Samples were prepared by drop-casting a suspension of diamond particles in 200 proof ethanol onto lacey carbon grids (Ted Pella, Inc.) and air-drying before transferring into the microscope.

### Micro Raman characterization
Micro Raman spectra were acquired at room temperature using a Nanophoton Raman 11 confocal Raman/PL microscope. Samples were prepared by drop-casting a suspension of diamond particles in 200 proof ethanol onto 1 cm x 1 cm pieces of MgO (100) (MTI Corporation). The spectrometer was calibrated using a Ne lamp spectrum as well as the Si $T_{2g}$ phonon mode at ~520.2 $cm^{-1}$. An excitation wavelength of 532 nm was used, and the laser power was limited to 0.5 mW to avoid sample heating. Spectra were collected in the range of 250-2200 $cm^{-1}$ with a spectral resolution of 1.6 $cm^{-1}$ and a peak position accuracy of 0.1 $cm^{-1}$ using a 600 groove/mm grating. A 50x objective lens was used to focus the laser down to a spot size of ~10 µm.

### Tip-enhanced Raman spectroscopy
Spatially resolved Raman spectroscopy measurements were performed using a Horiba XploRA PLUS Raman Microscope head-based tip-enhanced Raman spectroscopy (TERS) system allowing for co-localized and simultaneous TERS and atomic force microscopy (AFM) measurements with a lateral resolution of ~50 nm. In our measurements, the spatial resolution was limited by the photon collection volume needed to maximize signal-to-noise to 200 nm. Samples were prepared by drop-casting a suspension of BDD particles in 200 proof ethanol onto Si(100) with a 50 nm Au coating. Spectra were collected in near-field mode using a Au-coated conductive cantilever (Opus 160AC-GG) with an excitation wavelength of 532 nm and a spectral resolution of 1.4 $cm^{-1}$ using a 1200 groove/mm grating. Prior to all measurements, the spectrometer was calibrated using the Si $T_{2g}$ phonon mode at 520.2 $cm^{-1}$.

### STEM-VEELS
STEM-VEELS measurements were performed using a Thermo Scientific Themis Z microscope equipped with a dual-EELS spectrometer for simultaneous measurement of low-loss and core-loss EELS, and a monochromated field emission gun (FEG) with a spectral resolution of 100 meV. All measurements were conducted at an accelerating voltage of 300 kV in order to obtain a high spatial resolution of 1 nm$^2$. Samples were prepared the same as described above for TEM measurements. Raw spectra were first corrected using the built-in tools of the Gatan Digital Micrograph software to account for any drift in energy experienced during the scan by aligning the ZLP maximum at 0 eV. All spectra were normalized to the ZLP maximum to account for varying thickness across a given particle and from particle to particle.

**Photo-induced force microscopy (PiFM)**
PiFM measurements of BDD and undoped diamond were performed using a Molecular Vista One microscope. Samples were prepared by dropcasting suspensions of BDD and undoped diamond in ethanol onto glass slides with an IR-reflective coating to minimize interference from the substrate. The samples were subsequently baked in vacuum at 100 °C for 1 hour to eliminate any solvent residue. AFM cantilevers with conductive Pt-Ir coated tips (Molecular Vista, Inc.) were used to obtain spectra down to ~$10^2$ nm spatial resolution and 1 cm$^{-1}$ spectral resolution at a laser power of 2 mW. The spectrometer and all accompanying optics were calibrated using a standard polyethersulfone (PES) sample before each measurement.

**Infrared scattering scanning near-field optical microscopy (IR s-SNOM)**
s-SNOM measurements of BDD and undoped diamond were performed using a Neaspec IR-NeaSCOPE with a broadband MIR laser tunable from 700 to 2200 cm$^{-1}$ and a pseudoheterodyne detector to simultaneously acquire the near-field amplitude and phase. Samples were prepared by drop-casting suspensions of BDD and undoped diamond particles in ethanol onto clean pieces of Si(100), which also served as a reference for normalizing optical signals. In order to achieve a good optical contrast, a Pt-Ir$_5$ coated silicon AFM probe (NanoWorld Arrow™ NCPt), featuring a tip radius of 25 nm, was used for all measurements.[48] Scans were performed on 5x5 µm$^2$ area, with a pixel area of 100x100 pxl$^2$ using an integration time of 33 ms. The signal was demodulated at the second harmonic to obtain a purely optical response. Experimental drifts were monitored and corrected in the Gwyddion software,[49] using the mechanically and spectrally stable Si wafer as a reference.

**First principle calculations of energy-loss function and absorption coefficient**
The energy-loss function and absorption coefficient were obtained by evaluating all quantities in the outlined three-step procedure from first principles using density functional theory (DFT) as implemented in the PWscf code of the QuantumESPRESSO suite.[50] Norm-conserving pseudopotentials generated with the van Barth-Car approach were used to describe the electron-nuclei interaction. The electron-electron interaction was approximated on the level of the local density approximation in the parametrization of Perdew and Zunger. The electron wave functions were expanded in a plane wave basis using an energy cutoff of 90 Ry. We use the established lattice constant of a=3.523 Å for the diamond geometry.

The main challenge in calculating the Fermi energy shift and the contribution of valence band transitions to the macroscopic dielectric function is the evaluation of the integrals over the first BZ. As the effective Fermi level is close to the top of the valence band, a high resolution in **k**-space is needed, making it necessary to sample the BZ with a dense, uniform grid of points in order to achieve convergence. As the calculation of all energies and position operator matrix elements on such a dense **k**-point mesh is computationally demanding, we first obtained DFT band energies and wave functions on a coarse 8x8x8 **k**-point mesh. Subsequently, we used an interpolation scheme on the basis of maximally localized Wannier functions, as implemented in the Wannier90. The electronic band structure and the matrix elements of the velocity operator were then interpolated using the EPW code to a fine 200x200x200 **k**-point mesh. Finally, the position operator matrix elements were obtained via

$$\langle \mathbf{k}n|\hat{r}_i|\mathbf{k}m\rangle = -\frac{i\hbar}{m_e}\frac{\langle \mathbf{k}n|\hat{p}_i|\mathbf{k}m\rangle}{\varepsilon_{\mathbf{k}n}-\varepsilon_{\mathbf{k}m}} + \frac{\langle \mathbf{k}n|[\hat{H}_{non-local},\hat{r}_i]|\mathbf{k}m\rangle}{\varepsilon_{\mathbf{k}n}-\varepsilon_{\mathbf{k}m}} \text{ for } \varepsilon_{\mathbf{k}n} \neq \varepsilon_{\mathbf{k}m},$$

where $m_e$ denotes the electron mass and the $\hat{H}_{non-local}$ the non-local part of the pseudopotential.

**Ab initio calculations of phonon density of states**

First-principles calculations of the phonon density of states (PDOS) were performed with the same settings and approximations as the ground state calculations previously described for the macroscopic dielectric function. For the undoped PDOS, an 8x8x8 **k**-point mesh was used for integration of the electronic BZ. The phonon frequencies were obtained from density functional perturbation theory as implemented in the PHonon code of the QuantumESPRESSO suite. A threshold of $10^{-14}$ $Ry^3$ was assumed for the squared perturbed self-consistent potential in the self-consistency cycle. The phonon frequencies and dynamical matrices were first computed on a uniform 8x8x8 **q**-point mesh in the first phonon BZ, and subsequently Fourier interpolated to a denser 16x16x16 **q**-point mesh for the calculation of the PDOS, using the "crystal" acoustic sum rule to preserve exact translation invariance. The number of frequency points between the highest and lowest frequency was 1000 and the broadening was 4 $cm^{-1}$. To calculate the PDOS of BDD, , we constructed a 6x6x6 supercell of the diamond lattice, with 432 atoms in total, and replaced 1 carbon atom replaced by a boron atom, which corresponds to a boron concentration of ~2300 ppm. Any lower concentration would require a larger supercell which proved to be computationally prohibitively expensive. The atomic positions and lattice constant were both relaxed at convergence thresholds for the total energy of $10^{-7}$ Ry and for the total force components of $10^{-8}$ Ry/Bohr. As the system is formally metallic, a Fermi-Dirac smearing using a smearing of 0.002 Ry was applied. The value of the optimized lattice constant was found to be 21.141 Å, corresponding to an average unit cell lattice constant of 3.523 Å. Due to the size of the supercell, both the electronic and phonon BZ were sampled at only one point without interpolating the phonons.

## Data Availability

The data that support the findings of this study are available from the corresponding authors upon reasonable request.

## Acknowledgements

This material is based upon work supported by the National Science Foundation (NSF) under Grant No. DMR-1708742. N.M. acknowledges support from the Swedish Research Council (Grant No. 2021-05784), Kempestiftelserna (Grant No. JCK-3122), the Knut and Alice Wallenberg Foundation (Grant No. KAW 2023.0089), the Wenner-Gren Foundations (Grant No. UPD2022-0074), and the European Innovation Council (Grant No. 101046920). Electron microscopy was performed at the Center for Electron Microscopy and Analysis (CEMAS) at The Ohio State University and the authors thank Prof. David McComb and Dr. Robert Williams for helpful discussions and their technical help. Micro Raman spectroscopy, TERS, PiFM and s-SNOM were carried out at the Materials Research Laboratory Central Research Facilities at the University of Illinois Urbana-Champaign, which is partially supported by NSF through a Materials Research Science and Engineering Center under Grant No. DMR-2309037. The authors would also like to thank Prof. S. I. Bogdanov for insightful discussions.

## Authors Contributions Statement

R.M.S. and G.S. conceived the study. S.B. and J.B. carried out TEM and STEM-VEELS measurements and analyzed the results. S.B. performed micro-Raman spectroscopy, TERS, and PiFM measurements, and analyzed the results. S.B. and V.A. performed s-SNOM measurements and analyzed the results with input from A.L.L. J.B. carried out calculations using the Thomas-Fermi model with input from N.M. S.R. and A.H.T. carried out vibrational modeling of diamond with input from L.W. S.R. developed the first-principles model for the dielectric function and carried out calculations. S.B., J.B., G.S., S.R., and R.M.S. wrote the manuscript. All authors helped revise the manuscript to its final form.

# Supplementary Information

# Intervalence Plasmons in Boron-Doped Diamond

Souvik Bhattacharya,[1,*] Jonathan Boyd,[2,*] Sven Reichardt,[3,*], Valentin Allard,[4] Amir Hossein Talebi,[3] Nicolò Maccaferri,[5] Olga Shenderova,[6] Aude L. Lereu,[4] Ludger Wirtz,[3] Giuseppe Strangi,[2] and R. Mohan Sankaran[1]

[1]Department of Nuclear, Plasma, and Radiological Engineering, The Grainger College of Engineering, University of Illinois Urbana-Champaign, Champaign, IL, U.S.A.

[2]Department of Physics, Case Western Reserve University, Cleveland, OH, U.S.A.

[3]Department of Physics and Materials Science, University of Luxembourg, Luxembourg

[4]Aix Marseille Univ, CNRS, Centrale Méditérranée, Institut Fresnel, Marseille, France

[5]Department of Physics, Umeå University, Sweden

[6]Adamas Nanotechnologies, Raleigh, NC, U.S.A.

**Supplementary Note 1. Characterization of diamond samples by transmission electron microscopy.** Optical images of suspensions of BDD and undoped diamond crystals in ethanol are shown in Supplementary Fig. 1. The presence of boron is manifested visibly by the bluish appearance of the BDD suspension. Supplementary Figs. 2a and 2b show high-resolution transmission electron microscopy (HR-TEM) images of the BDD and undoped diamond samples, respectively. The samples were found to consist of nearly identically-sized particles on the order of ~1 μm (inset, Supplementary Figs. 1a and 1b). The lattice fringes confirm the crystalline nature of these particles. We note that despite the addition of boron, there is no observable loss of crystallinity in the boron-doped samples. Moreover, to preserve the crystallinity of these microdiamonds, no special TEM sample preparation or post-treatment was performed on either sample.

Supplementary Fig. 1c shows the core-loss region of the electron energy loss (EEL) spectra obtained from BDD and undoped diamond, respectively. We observe a sharp peak ca. 292 eV for both samples, which is ascribed to σ* transitions corresponding to $sp^3$-hybridized carbon atoms indicative of crystalline diamond, and a smaller peak at ca. 284 eV, which is ascribed to π* transitions corresponding to $sp^2$-hybridized carbon atoms indicative of graphitic carbon. The π* peak is small compared to the σ* peak, confirming a high $sp^3/sp^2$ ratio in these samples. Despite the stark contrast in color of these samples, we did not observe a B K-edge signal in the core-EELS region which would have occurred from ca. 190 eV to ca. 240 eV.[1]

In order to try to observe the B K-edge, which would allow us to co-locally study the variation of VEELS with the presence of boron, we analyzed the core EELS spectra further by integrating over different areas to accumulate the signal (see Supplementary Fig. 3). In all cases, no B K edge signal was observed. The lack of a signal is consistent with the boron concentration in our samples, which is <1000 ppm (see Supplementary Note 2), being below the detection limit of core-EELS generally at or above a few at. %.[1,2]

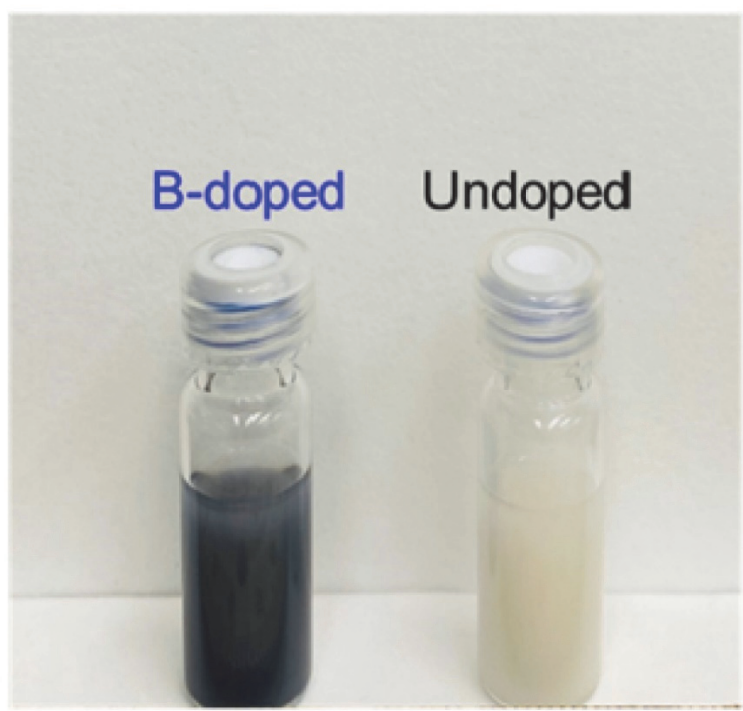

**Supplementary Fig. 1: Photographs of diamond samples used in this work**

Photographs of BDD (left) and undoped diamond (right) crystals synthesized using high-pressure high-temperature (HPHT) dispersed in ethanol. The BDD sample appears deep blue due to the presence of the boron dopant.

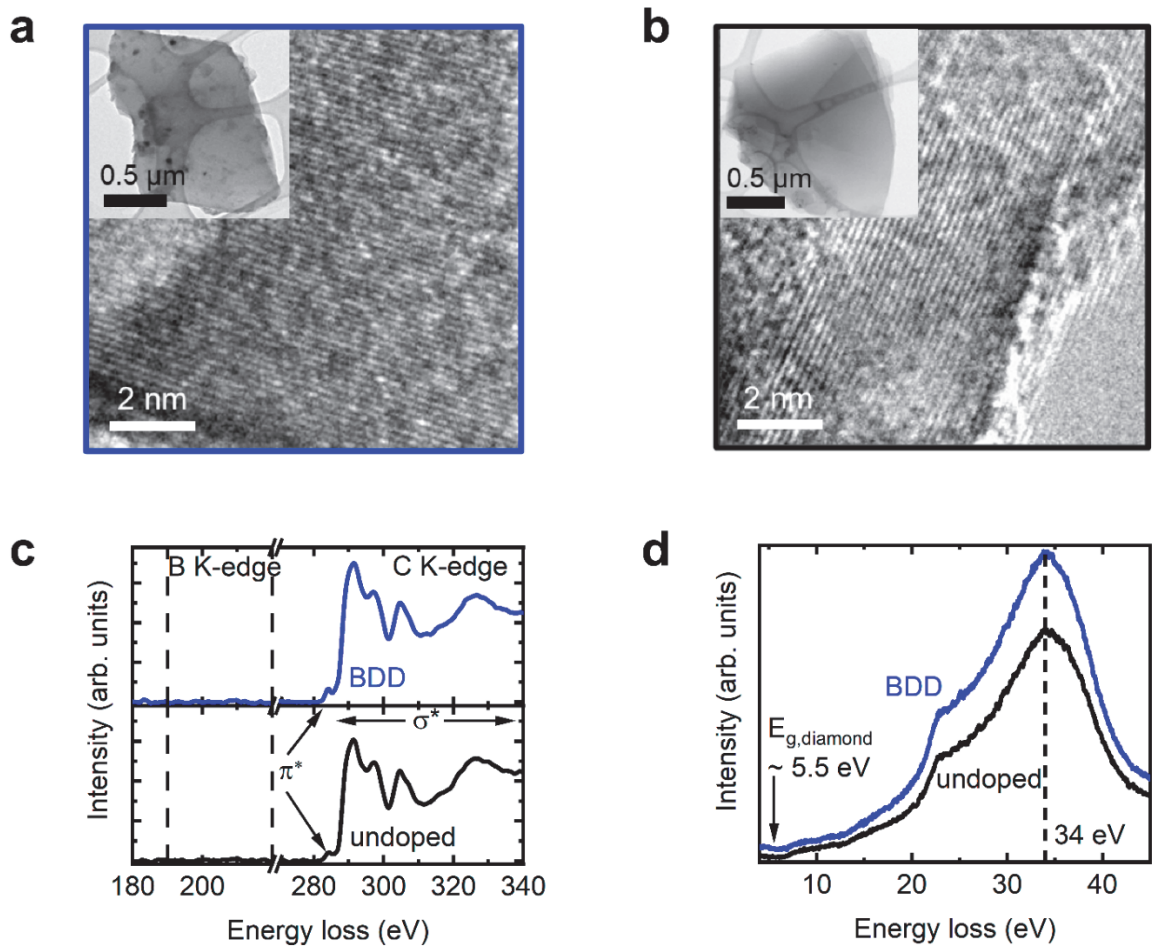

**Supplementary Fig. 2: Characterization of BDD and undoped diamond samples**

**a-b**, Representative HRTEM images of BDD, **(a)** and undoped diamond, **(b)** particles showing lattice fringes. The insets show that the diamond samples consist of identically sized particles. **c**, Core-EELS spectra acquired from BDD and undoped diamond samples. While a clear C K-edge signal was observed in both samples showing a high $sp^3/sp^2$ ratio, no B K-edge signal was observed from the BDD sample. **d**, EELS spectra acquired from BDD and undoped diamond showing the volume plasmon resonance at ca. 34 eV, proportional to the valence electron density of diamond. The onset of this signal occurs at ca. 5.5 eV which coincides with the bandgap of diamond.

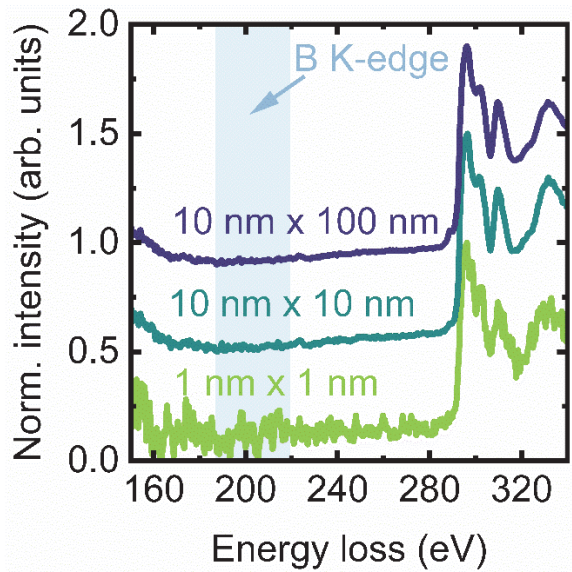

**Supplementary Fig. 3: Core-loss EELS analysis of the B K-edge signal**

Core-loss STEM-EEL spectra of BDD collected with 1 nm x 1 nm spatial resolution and integrated over different areas. The spectra were corrected for background by a power law fit. The absence of a discernable B K-edge signal in all cases is due to the boron concentration being below the detection limit of EELS.

**Supplementary Note 2. Modeling the Fano lineshape measured in the micro Raman spectra of BDD to estimate the average free charge carrier concentration.** The average hole density in BDD was estimated by analysis of the Fano interference observed in micro Raman spectra of semiconductor materials, which occurs as a result of the interaction of phonons with a continuum of electronic excitations such as those opened up by hole doping in BDD. This method is non-destructive and sensitive to dopant concentrations down to 100 ppm.[3] The comprehensive overall procedure developed by Mortet et al. was used to model the asymmetric line shape by a combination of two discrete excitations, the zone-center phonon (ZCP) and the phonon-density-of-states (PDoS) maximum, and a continuum termed as the electronic Raman scattering (ERS) background, the latter of which is a direct consequence of the holes created by the introduction of boron.[4] The model was fit to the experimentally-obtained spectra using least squares regression ($R^2$=0.968), and the converged values of the fit parameters such as the spectral position shifts and FWHMs of the Fano-modified ZCP (1327 cm$^{-1}$), and the PDoS maximum (1295 cm$^{-1}$), were found to correspond to a boron doping level below the semiconductor-metal transition ([B] $\leq 4 \times 10^{20}$ cm$^{-3}$). The parameter of interest in this framework is the asymmetry coefficient, $q$, which is defined as the ratio of the transition probability of the discrete excitation to that of the continuum excitations, and whose converged absolute value is inversely correlated with the charge carrier density.[5,6] By comparing $q_{ZCP}$ (= -164) and $q_{PDoS}$ (= -0.41) obtained from our analysis with reported values measured by Hall measurements of crystalline BDD films,[4] we estimated the average hole density of our BDD sample to be $\sim 5 \times 10^{19}$ cm$^{-3}$. It is noteworthy that this value is an average over several BDD particles and does not account for doping inhomogeneity across individual BDD particles.

**Supplementary Note 3. Tip-enhanced Raman spectroscopy (TERS) to probe the spatial variation of boron doping at the nanoscale.** As detailed in Supplementary Note 2, micro Raman characterization of BDD shows a Fano resonance that can be modeled to obtain the boron concentration. However, because the laser spot size in micro Raman spectroscopy is ~10 µm and the particles are ~100s of nm in size, the boron concentration that is obtained is an average over multiple particles. To obtain the boron concentration with smaller spatial resolution, we performed TERS. In TERS, nanometer scale resolution is possible with the main limitation being sufficient signal. Supplementary Fig. 4 shows TER spectra collected from a representative BDD particle over 3 different scales: ~1 µm, 400 nm, and 200 nm, respectively. As the spatial area decreases, the Fano-type lineshape becomes more difficult to observe and below 200 nm was not clear. Nonetheless, we were able to improve the spatial resolution as compared to micro Raman and map single particles. The TER spectra were modeled by the same procedure discussed previously (see Supplementary Note 2). At 1 µm, the asymmetry factor of the PDoS contribution ($q_{PDoS}$) obtained post-modeling ($R^2 = 0.947$) was found to be -1.15 while $q_{ZCP}$ was a large negative value (Supplementary Fig. 4a). The boron doping level for these parameter values corresponds to roughly ~800 ppm based on similar modeling in literature, which represents an average over a single particle and is very close to the doping level obtained by micro Raman spectroscopy.[4] Starting at 400 nm, variations were observed from different regions (Supplementary Fig. 4b). The $q_{PDoS}$ from two regions were found to be -2.3 and 0.2, respectively, while $q_{ZCP}$ were large negative values ($R^2 = 0.956$ and 0.937, respectively), corresponding to boron doping levels of ~200-400 ppm and ~1000 ppm, respectively. For 200 nm (Supplementary Fig. 4c), the $q_{PDoS}$ from two regions (2 and 3) were calculated to be roughly similar at ~0.4 ($R^2 = 0.965$ and 0.951 respectively), corresponding to a boron doping level of ~200-400 ppm, and two regions (1 and 4) were below the limit of fitting the Fano-line-shape and therefore, less than ~200 ppm. Overall, TERS confirms that the boron concentration varies within single particles over a relatively wide range, from less than 200 ppm to ~1000 ppm.

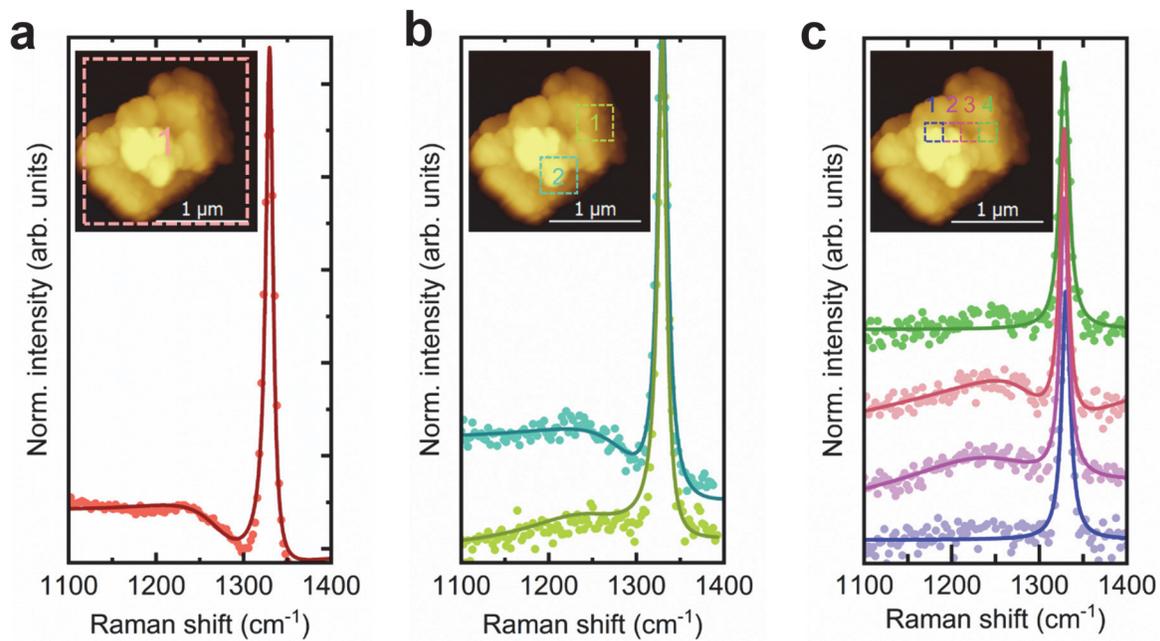

**Supplementary Fig. 4: TERS characterization and analysis of BDD**

Experimental TER spectra (scatter points) and corresponding model fits of the Fano lineshape (solid line) for a representative BDD particle at spatial resolutions of **a**, 1 µm. **b**, 400 nm, and **c**, 200 nm. For 400 and 200 nm, TER spectra were collected from different regions of the particle. Inset shows an AFM image of the BDD particle and locations corresponding to TER spectra. Variation of the Fano-type lineshape indicates heterogeneity of boron doping within a single particle.

**Supplementary Note 4. Isolation and analysis of inelastic VEELS signal from BDD and undoped diamond particles**. Methods to fit and subtract the elastic ZLP component in order to isolate the inelastic component from raw VEEL spectra remain challenging and are an active area of research.[7,8] As shown in Supplementary Fig. 5, two different methods were used here to analyze the inelastic VEEL signal. In the first method, which we term the 'spectral fitting' method, the ZLP was modeled by a vacuum VEEL spectrum, and the inelastic signal was fit to a Voigt function to generally account for any Gaussian and Lorentzian contributions (see Supplementary Fig. 5c and 5d). To independently confirm this approach, a more conventional procedure to separate the inelastic signal by subtracting the ZLP from VEEL spectra was also explored (see Supplementary Fig. 5e and 5f).[8] Both methods produced identical spectral position and overall shape of the inelastic signal. However, the subtraction method produced non-physical features such as negative signals. Therefore, the fitting method was chosen for analysis of all VEEL spectra. Supplementary Figs. 6 and 7 show VEELS spectra obtained from BDD and undoped diamond samples, respectively, where the inelastic VEEL signal unique to BDD manifests itself as a shoulder or asymmetry on the zero-loss peak (ZLP).

The spatial variation of the inelastic signal across particle areas was studied by collecting spectra at each 1 nm x 1 nm pixel of the STEM images and applying the same fitting method. Color maps were constructed for different spectral energies by integrating the inelastic VEEL signal within a narrow energy slice centered around the energy value of interest. Color maps corresponding to 0 eV accounted for the attenuation of the ZLP resulting from variation in the thickness of the sample. The intensities were normalized by subtracting this contribution of the sample thickness in order to only show variation from inelastic electron scattering processes (Supplementary Figs. 9 and 11).

We further analyzed the variation of the inelastic VEELS signal with thickness by quantifying the thickness of the particles using a previously reported log-ratio method.[9] Assuming an inelastic mean free path of 150 nm for 150 keV electrons in diamond[10], the calculations show that the thinnest regions near the edge of the particles correspond to ~2-3 nm in thickness, while the thicker regions into the body of the particles correspond to ~15-20 nm in thickness. Despite this relatively large variation in thickness, the measured inelastic VEEL signal at ca. 0.15 eV shows a weak correlation, indicating that the signal is largely independent of the sample thickness.

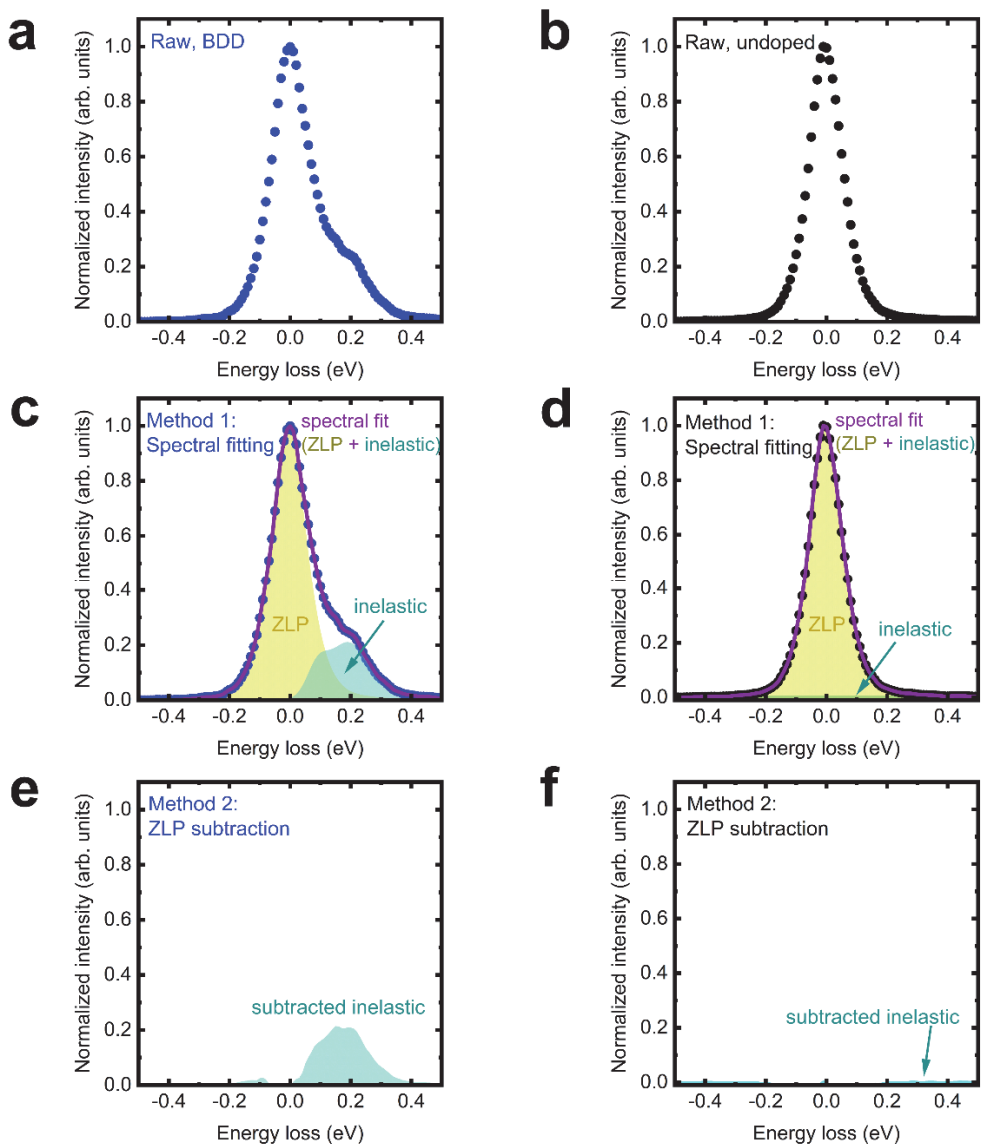

**Supplementary Fig. 5: Isolation and analysis of the inelastic component of VEEL spectra**

Analysis of VEEL spectra collected from **a**, BDD and **b**, undoped diamond by two methods: **c**, **d**, spectral fitting, and **e**, **f**, ZLP subtraction. Both methods independently confirm identical inelastic VEEL signatures present only in BDD.

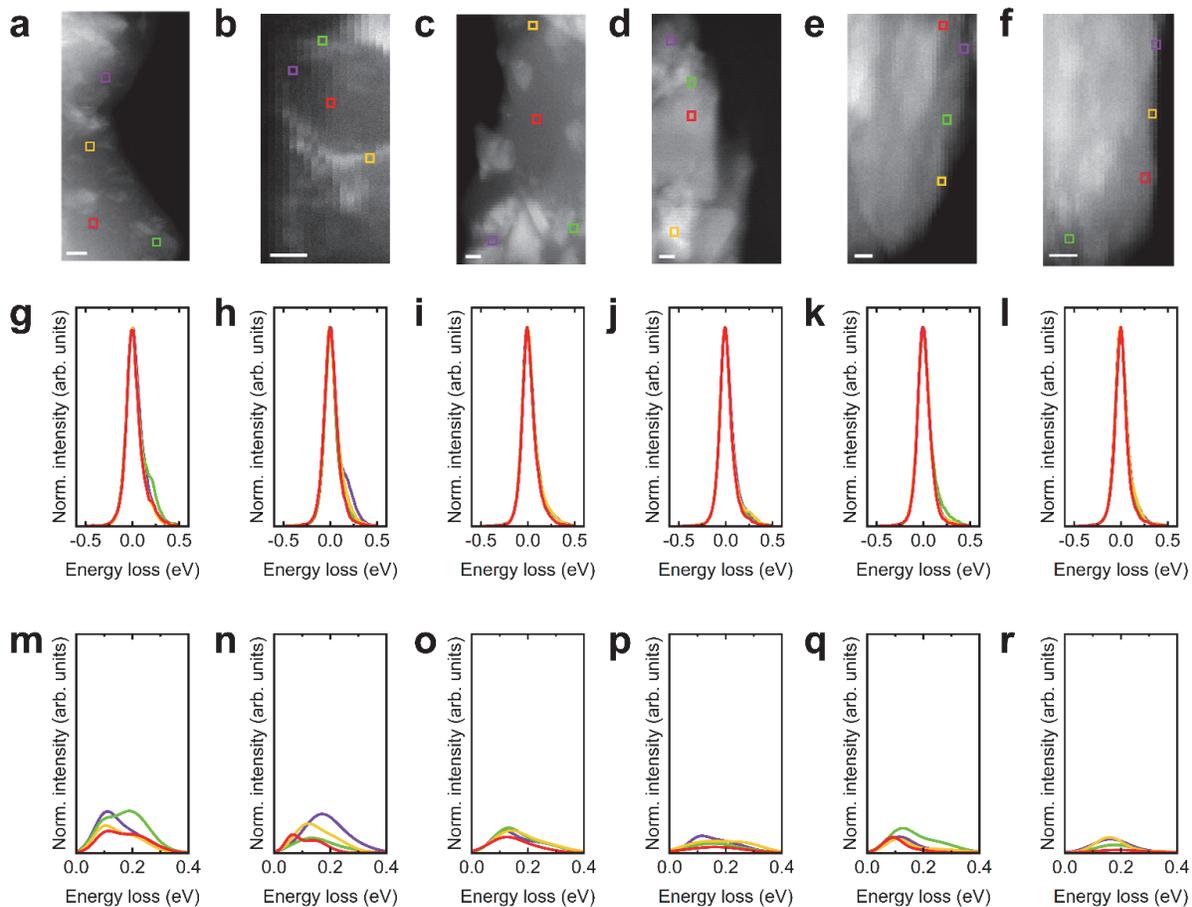

**Supplementary Fig. 6: Experimental VEEL spectral analysis of different BDD particles**

**a-f**, Annular dark field (ADF) images of sections of different BDD particles. All scale bars correspond to 10 nm. **g-l**, Raw VEEL spectra taken at 4 1 nm x 1 nm spots from each BDD particle shown in **(a)-(f)**. **m-r**, Residual inelastic VEEL signal isolated after separation from the ZLP from the raw spectra shown in **(g)-(l)**.

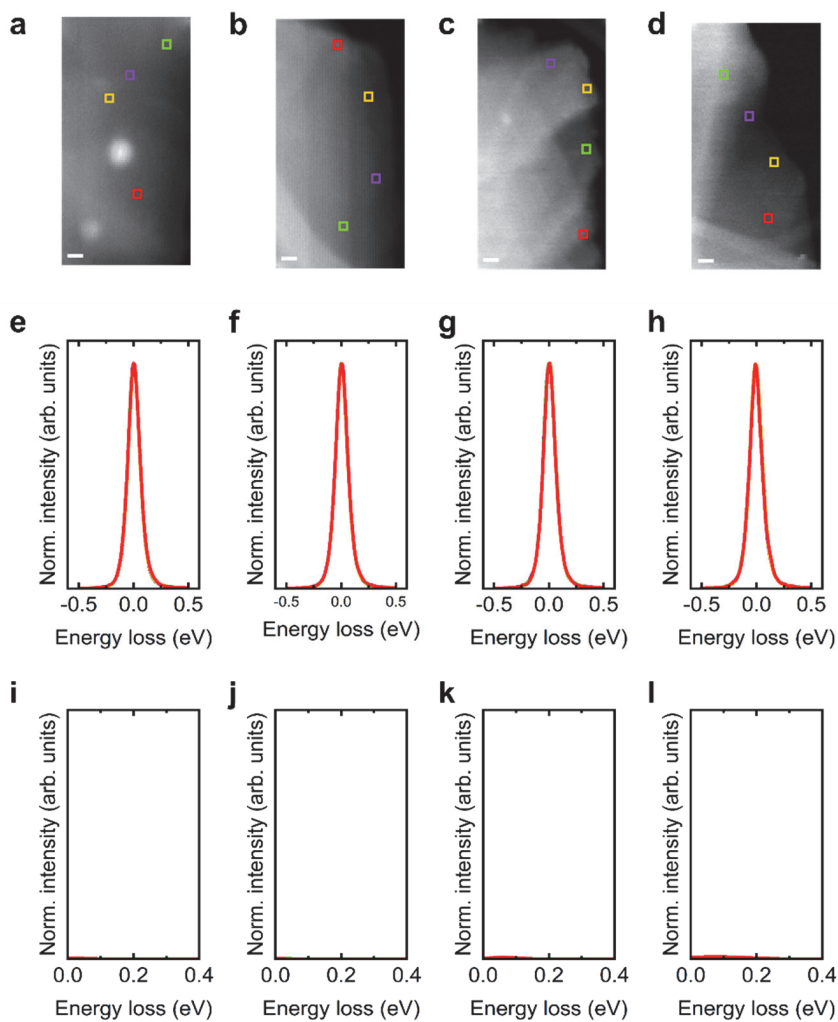

**Supplementary Fig. 7: Experimental VEEL spectral analysis of different undoped diamond particles**

**a-d**, Annular dark field (ADF) images of sections of different undoped diamond particles. All scale bars correspond to 10 nm. **e-h**, Raw VEEL spectra taken at 4 1 nm x 1 nm spots from each undoped diamond particle shown in **(a)-(d)**. **i-l**, Residual inelastic VEEL signal isolated after separation from the ZLP from the raw spectra shown in **(e)-(h)**.

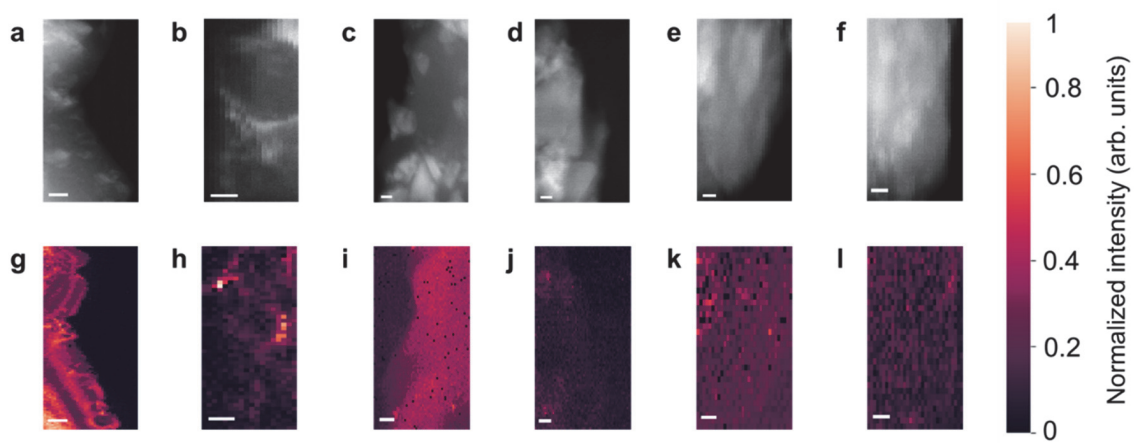

**Supplementary Fig. 8: VEELS intensity color maps for BDD**

**a-f**, Annular dark field (ADF) images of different BDD particles. **g-l**, Color maps showing spatial variation of the normalized inelastic VEEL signal at 0.15 eV energy for each particle shown in **(a)-(f)**. All scale bars correspond to 10 nm and the color bar corresponds to the normalized inelastic VEEL intensity on a scale of 0 to 1 in arbitrary units.

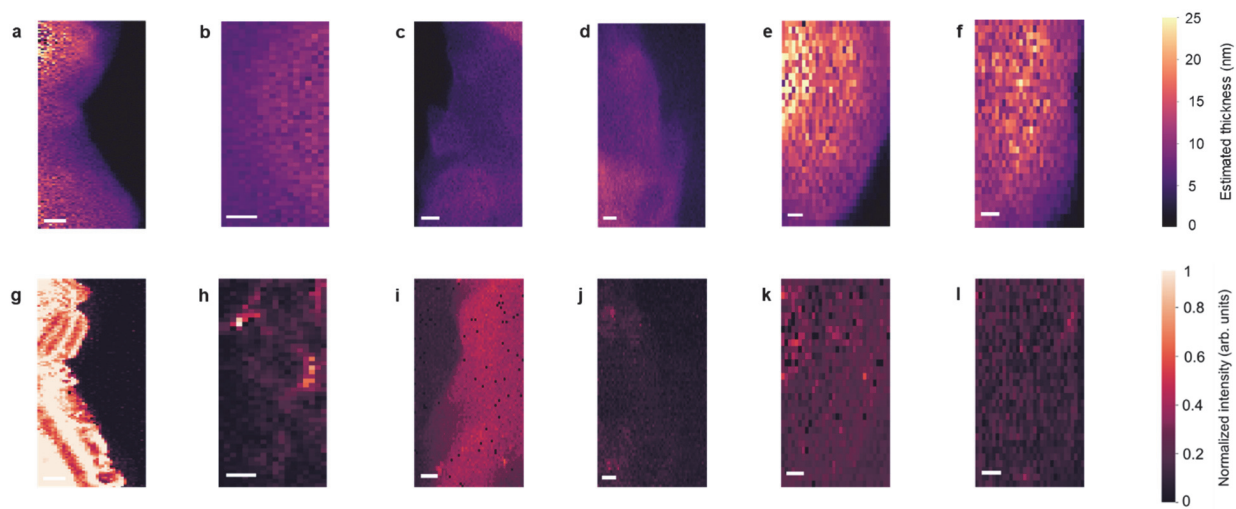

**Supplementary Fig. 9: VEELS intensity correlation with sample thickness**

**a-f**, False color maps showing the variation in particle thickness calculated by the log-ratio method for different BDD particles assuming an inelastic mean free path of 150 nm for 300 keV electrons. **g-l**, Color maps showing spatial variation of the normalized inelastic VEEL signal at 0.15 eV energy for each particle shown in **(a)-(f)**. All scale bars correspond to 10 nm and the color bar corresponds to the normalized inelastic VEEL intensity on a scale of 0 to 1 in arbitrary units.

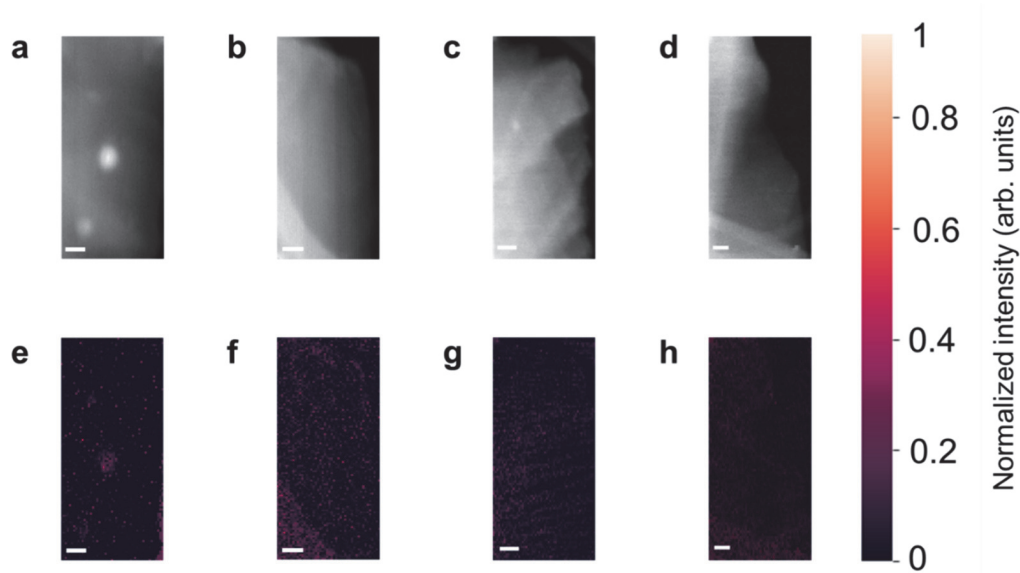

**Supplementary Fig. 10: VEELS intensity color maps for undoped diamond**

**a-d**, Annular dark field (ADF) images of different undoped diamond particles. **e-h**, Color maps showing spatial variation of the normalized inelastic VEEL signal at 0.15 eV energy for each particle shown in **(a)-(d)**. All scale bars correspond to 10 nm and the color bar corresponds to the normalized inelastic VEEL intensity on a scale of 0 to 1 in arbitrary units.

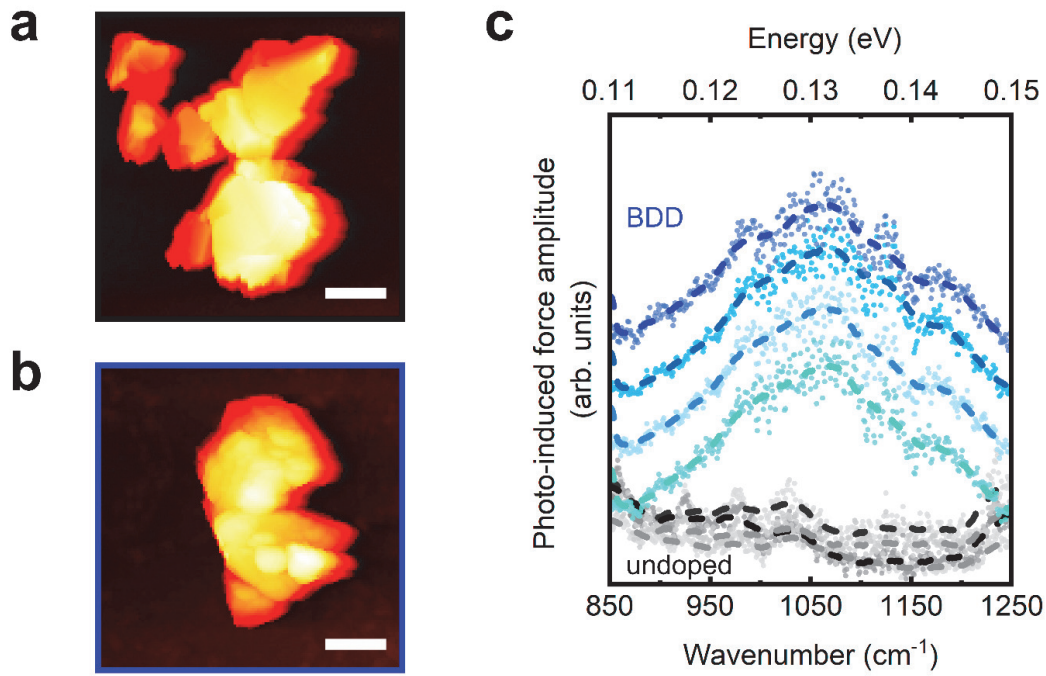

**Supplementary Fig. 11: Photo-induced force microscopy (PiFM) characterization of BDD and undoped diamond**

**a-b**, Topographical scans of representative undoped diamond **(a)** and BDD **(b)** particles. The scale bar represents 1 $\mu$m. **c**, PiFM spectra collected in the mid-IR regime from the undoped diamond (gray plots) and BDD (blue plots, offset for clarity) shown in **(a)** and **(b)**. A peak around 0.13 eV is observed only for BDD which is correlated with absorbance.

**Supplementary Note 5. s-SNOM analysis of BDD and undoped diamond.** Scattering-scanning nearfield optical microscopy (s-SNOM) is a nanoscale characterization technique in which an incident electromagnetic field is focused beneath the tip of a metalized AFM probe to generate spatially-resolved images of the amplitude and phase shift of the scattered field.[11-14] In this study, s-SNOM was conducted by imaging BDD and undoped diamond crystals on a Si substrate, with a laser energy of 1048 cm$^{-1}$ consistent with the resonance observed by VEELS and PiFM (see Methods for experimental details). Since the scattered field, $E_s$, and the incident field, $E_i$, are proportional, i.e., $E_s \propto \sigma E_i$, the scattering coefficient, $\sigma$, can be used to obtain the sample's optical response to the incident excitation (phase and amplitude shift). The nearfield contrast, $\eta_n$, at the $n^{th}$ demodulation harmonic is then computed as:[15]

$$\eta_n = \frac{\sigma_n}{\sigma_{n,ref}} = \frac{a_n}{a_{n,ref}} e^{i(\varphi_n - \varphi_{n,ref})}$$

where, $\eta_n$, $\sigma_n$ and $\sigma_{n,ref}$ are complex quantities and $a_n$ and $\varphi_n$ are the optical amplitude and phase shifts induced by the interaction of the sample with the near-field infrared light. Given the optical properties of Si[16] used in our experiments, the phase-shift of the reference area is set to 0 (light blue) and the amplitude to 1 (yellow), as shown in Fig. 4. Any changes in amplitude and phase are then directly linked to variations in the optical near field interaction induced by the respective samples.

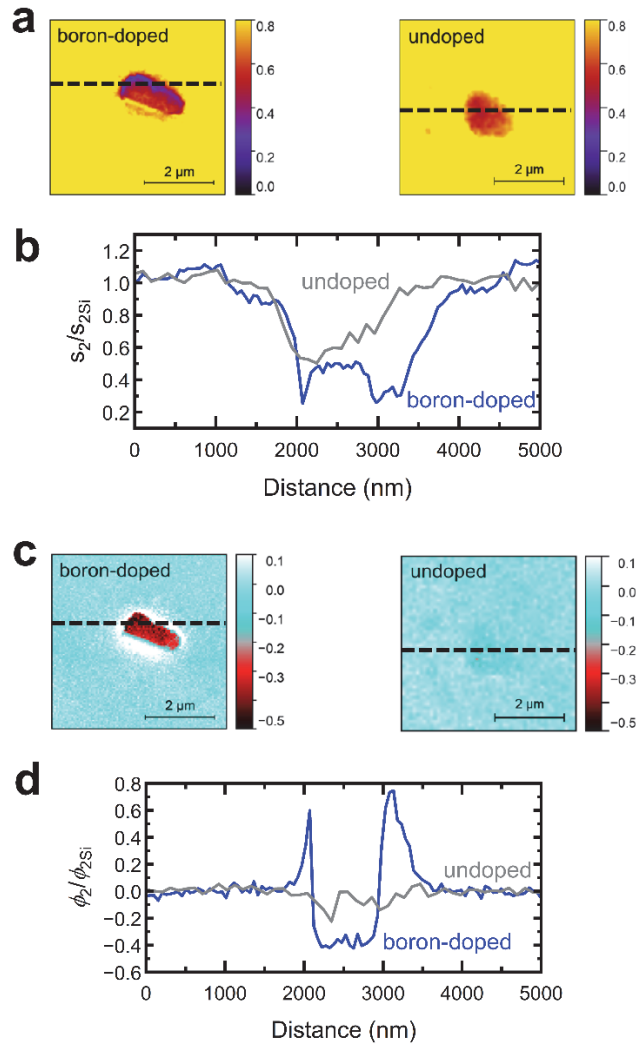

**Supplementary Fig. 12: s-SNOM amplitude and phase profiles for BDD and undoped diamond.**

**a,b**, Normalized amplitude images of BDD and undoped diamond **(a)** and corresponding profiles **(b)** extracted along the dashed lines showing similar variations in amplitude. **c,d**, Normalized phase images of BDD and undoped diamond **(c)** and corresponding profiles (d) along the dashed lines showing a phase contrast unique to BDD **(d)**.

**Supplementary Note 6. Thomas-Fermi model analysis of energy-loss function.** A Drude-like model analysis of the energy-loss function was performed by applying the Thomas-Fermi model that has previously been used to predict infrared plasmons in heavily doped p-type and n-type silicon.[17] In this framework, the macroscopic dielectric permittivity of the material, $\varepsilon_M(\omega)$, is given by:

$$\varepsilon_M(\omega) = \varepsilon_0 \left( 1 + \frac{i\omega_p^2 \tau}{\omega(1 - i\omega\tau)} \right)$$

where $\varepsilon_0$ is the relative static dielectric constant for diamond, $\omega_p$ is the plasma frequency, and $\tau$ is the collision time. The plasma frequency is further related to the hole density, $n_H$ by:

$$\omega_p^2 = \frac{4\pi e^2 n_H}{\varepsilon_0 m^*}$$

where $e$ denotes the elementary charge in Gauss units and $m^*$ is the effective mass of the charge carrier, which are holes in the case of BDD, and is known to be dependent on contributions from both light holes (LH) and heavy holes (HH) in the valence band[18] The plasma frequency was calculated as a function of hole density using a single effective mass that accounts for contributions from the LH and HH bands[19], and further, by constructing the total macroscopic dielectric function, the energy-loss function was simulated for a range of hole densities (Supplementary Fig. 12).

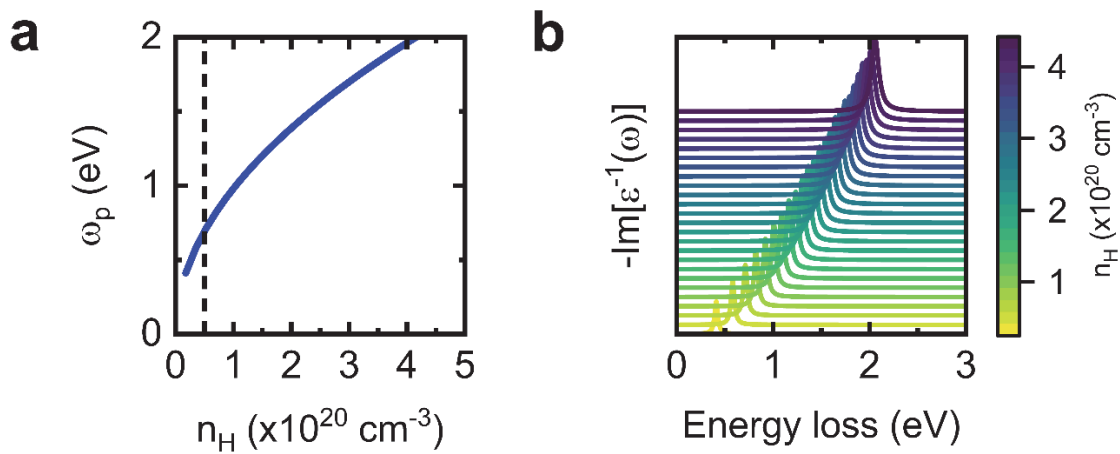

**Supplementary Fig. 13: Thomas-Fermi model analysis of plasmonic behavior in BDD**

**a**, Calculated plasma frequency, $\omega_p$, as a function of hole density, $n_H$, in BDD. The dashed line corresponds to the average hole density of our BDD sample which according to this framework, predicts a plasma frequency at ca. 0.7 eV (Supplementary Note 5). **b**, Simulated loss spectra as a function of hole density.

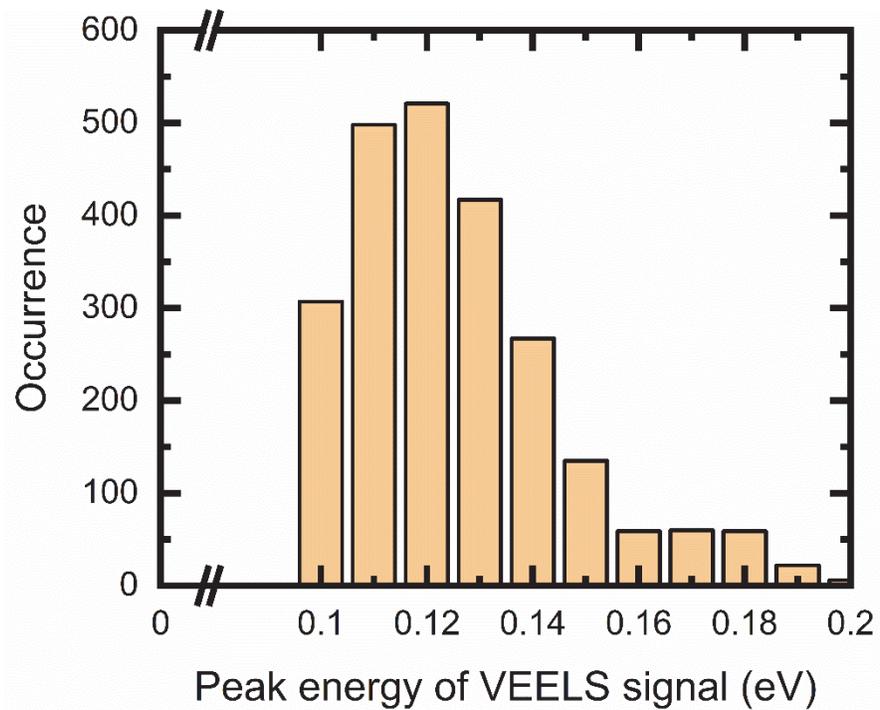

**Supplementary Fig. 14: Histogram of VEELS peak energies**

Statistical analysis of the peak energies measured by VEELS from 1 nm x 1 nm pixels over a 134 nm x 39 nm section of a representative BDD particle shown in Fig. 2c of the manuscript. The results show a distribution of energies between ~0.08 and 0.19 eV, with the highest occurrence at 0.12 eV.

**Supplementary Note 7. Ab initio calculations of energy-loss function and absorption coefficient.** First principles calculations of the energy-loss function, defined as $Im(-1/\varepsilon_M)$, and absorption coefficient, defined as $2\frac{\omega}{c} Im(\sqrt{\varepsilon_M(\omega)})$, where $\varepsilon_M$ is the total macroscopic dielectric function, were performed using ab initio methods following a three-step procedure: 1) defining an effective Fermi level as a function of the hole density, 2) relating the macroscopic dielectric function to the electronic transitions, and 3) constructing the total macroscopic dielectric function.

*Effective Fermi level for BDD*

The valence band structure of BDD was modeled by applying an effective Fermi level to the valence band structure of intrinsic diamond. The Fermi energy, $\varepsilon_F$, was calculated as a function of the hole density, $n_H$, by integrating the sum of the occupancies, $f_{\mathbf{k}n}$, of all electronic valence band states $|\mathbf{k}n\rangle$ over the first Brillouin zone (BZ):

$$8 - V_{UC} n_H = 2 \int_{BZ} \frac{d^3k}{V_{BZ}} \sum_{n \in Valence\ bands} f_{\mathbf{k}n}(\varepsilon_F)$$

where $V_{UC}$ and $V_{BZ}$ denote the volumes of the unit cell of diamond and the first BZ, respectively, the prefactor of two is related to the spin degeneracy of the bands, and at zero temperature the occupancies are given in terms of the Heaviside step function:

$$f_{\mathbf{k}n}(\varepsilon_F) = \Theta(\varepsilon_F - \varepsilon_{\mathbf{k}n}).$$

We note that the integral and sum on the right-hand side yield the total number of valence electrons per unit cell of diamond, which is equal to 8 for undoped diamond and is decreased by the number of holes in one unit cell, given by $V_{UC} n_H$. The equation above is a non-linear equation in $\varepsilon_F$, which we solve numerically using a golden section search. The calculation uses the experimentally-determined hole density and as such is independent of the exact energy level of the boron acceptor state, which could be modified by Coulomb interactions and is not easily captured in first-principles calculations.

*Macroscopic dielectric function for BDD*

On the level of non-interacting, single-particle transitions, the relation between the electronic transitions and their contribution to the macroscopic dielectric function is given by their independent-particle dielectric response function:

$$\varepsilon_M(\omega)|_{Valence}$$

$$= -\frac{16\pi e^2}{V_{UC}} \int_{BZ} \frac{d^3k}{V_{BZ}} \sum_{\substack{n,m \in Valence \\ bands}} \left\{ [1 - f_{\mathbf{k}n}(\varepsilon_F)] f_{\mathbf{k}m}(\varepsilon_F) |\langle \mathbf{k}n|\hat{r}_x|\mathbf{k}m\rangle|^2 \frac{\varepsilon_{\mathbf{k}n} - \varepsilon_{\mathbf{k}m} + i\gamma}{(\hbar\omega)^2 - (\varepsilon_{\mathbf{k}n} - \varepsilon_{\mathbf{k}m} + i\gamma)^2} \right\}$$

where $e$ is the elementary electronic charge in Gaussian units, $V_{UC}$ denotes the volume of one unit cell of diamond, $\hat{r}_x$ is the x-component of the position operator (the y- and z-components yield the same result, due to the cubic symmetry of the crystal), and $\gamma$ is a broadening constant that accounts for the finite lifetime of electronic transitions. The sum runs over all valence band combinations, $n, m$, but is constrained by the occupancy factors to pairs of occupied and unoccupied states, which for BDD were given by the Fermi energy shift. Since we consider the macroscopic dielectric function, only zero-momentum excitations, i.e., "vertical" electronic transitions, contribute.

*Total macroscopic dielectric function*

We also considered contributions of electronic transitions in intrinsic diamond to the macroscopic dielectric function of BDD, which stem from interband transitions between the valence and conduction bands. The total macroscopic dielectric function was then defined as:

$$\varepsilon_M(\omega) = \varepsilon_o + Re[\varepsilon_M|_{Valence}(\omega)] + i\, Im[\varepsilon_M|_{Valence}(\omega)]$$

where $\varepsilon_o$ is the static dielectric constant of diamond. Since we focus on the low-frequency region <0.3 eV, which is much smaller than the band gap of diamond, the static approximation for the contribution of interband transitions well-justified.

*Direct calculation of intervalence plasmon properties*

The frequency, $\omega_{pl}$, and dampening constant, $\gamma_{pl}$, of the intervalence plasmon can directly be obtained from the conditions:

$$Re[\varepsilon_M(\omega_{pl})] = 0$$

and

$$\gamma_{pl} = \frac{2\, Im[\epsilon(\omega_{pl})]}{\left.\frac{dRe[\varepsilon(\omega)]}{d\omega}\right|_{\omega=\omega_{pl}}},$$

with the latter expression obtained by a Taylor expansion of the real part of the dielectric function appearing in the denominator of the loss function. For the numerical evaluation of the derivative of the real part of the dielectric function at the plasmon frequency, we first convolved the real part of the dielectric function on a finite set of frequency points with a Savitzky-Golay filter to smoothen numerical artifacts related to the finite **k**-point sampling in the BZ integral appearing in the dielectric function.

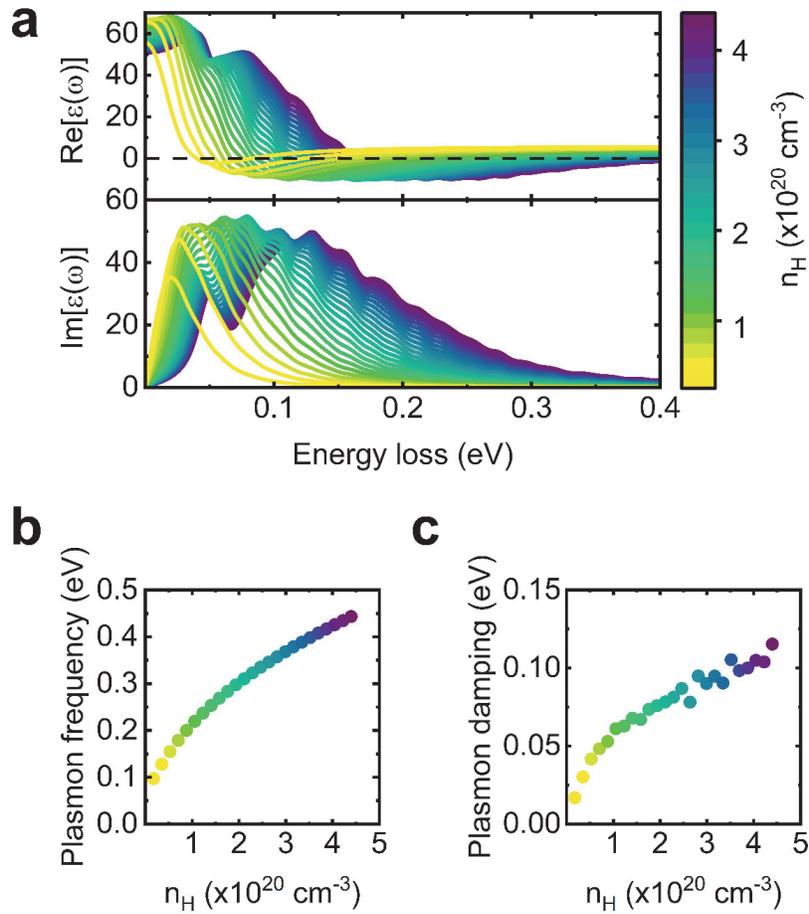

**Supplementary Fig. 15: Spectral analysis of the ab-initio dielectric function for BDD**

**a**, The real (top) and imaginary (bottom) parts of the simulated dielectric function for the different hole concentrations shown in Fig. 3. The real part exhibits a positive-to-negative crossover which meets the definition of a plasmon resonance. **b**, The intervalence plasmon frequency as a function of hole density shown in (a). **c**, The plasmon damping as a function of hole density, which is inversely related to the plasmon lifetime (Supplementary Note 5).